\documentclass[aps,pra,twocolumn,nofootinbib,superscriptaddress,floatfix]{revtex4-2}

\usepackage[T1,T2A]{fontenc}
\usepackage[utf8]{inputenc}
\usepackage[russian,english]{babel}

\usepackage{amsmath}
\usepackage{amssymb}
\usepackage{amsfonts}
\usepackage{amsthm}
\usepackage{bm}

\usepackage{textcase}

\usepackage{cancel}

\usepackage[colorlinks=true,linkcolor=blue,citecolor=blue,urlcolor=blue]{hyperref}

\theoremstyle{definition}

\newtheorem{remark}{Remark}

\newcommand\Tr{\operatorname{Tr}}

\renewcommand\Re{\operatorname{Re}}
\renewcommand\Im{\operatorname{Im}}

\newcommand{\braket}[1]{\left\langle #1\right\rangle}
\newcommand{\mf}[1]{\mathfrak{#1}}

\def\al {\alpha}
\def\be {\beta}

\def\de {\delta}
\def\DE {\Delta}

\def\ga {\gamma}
\def\GA {\Gamma}

\def\la {\lambda}

\def\om {\omega}
\def\OM {\Omega}

\def\pr {\partial}
\def\sg {\sigma}

\def\ve {\varepsilon}

\usepackage{graphicx}


\begin{document}
	
	\title{Joint Dynamics of Spins and Frequency-Resolved Bosonic Occupations: A Projection-Operator Approach}

	\author{K.Sh. Meretukov}
	\email{meretukov.khazret@gmail.com}	
	\author{A.E. Teretenkov}%
	\email{taemsu@mail.ru}
	\affiliation{Department of Mathematical Methods for Quantum Technologies, Steklov Mathematical Institute of Russian Academy of Sciences, ul. Gubkina 8, Moscow 119991, Russia
	}%

	\date{\today}
	
	\begin{abstract}
		 We derive weak-coupling kinetic equations for frequency-resolved bosonic mode occupations in the spin--boson model. Rather than integrating out the bath completely, we retain these mode occupations as independent dynamical variables whose equations are coupled to those for the spin observables.
		A projection-operator construction closes the dynamics on a product-spin ansatz and an undisplaced Gaussian environment with a diagonal covariance matrix. The equations are obtained to second order in the interaction parameter $\lambda$ for an arbitrary number of spins within this factorized closure, and are solved numerically for a single spin. A fixed-truncation  calculation using the hierarchical equations of motion  provides a benchmark for the spin dynamics with the Drude--Lorentz spectral density. The numerical results reveal  a redistribution of bosonic  mode occupations concentrated near the spin transition frequency $\Omega=\sqrt{\Delta^2+h^2}$. We also  examine the response of the coupled spin--boson system to structured spectra, nonthermal initial occupations, and slow modulation of the coupling. The calculated mode occupations also allow changes in the bath-side occupation and energy densities $p(\omega)\delta\braket{N_\omega}(t)$ and $\omega p(\omega)\delta\braket{N_\omega}(t)$ to be obtained without requiring an exponential decomposition of the bath correlation function. Its scope is restricted to weak coupling, diagonal Gaussian bath states, and, for time-dependent couplings, the adiabatic regime.
	\end{abstract}
	
	\maketitle
	
	\section{Introduction}
	Open quantum systems exchange energy and information with their surroundings, producing dissipation and decoherence \cite{Breuer2002, Weiss2012, Zurek2003}. Most reduced descriptions eliminate the environmental degrees of freedom and therefore do not retain where energy is deposited in the bath. Here we instead retain the frequency-resolved bosonic occupations together with the spin observables. This bath-side information is relevant when individual modes can be prepared or probed, as in engineered trapped-ion, microwave, and phononic environments.

	The central idea in studying such systems is to consider the system and its environment as a single composite system, which is closed by construction \cite{Breuer2002}. Standard evolution equations can then be written for this enlarged system. When the characteristic environmental correlation time $\tau_B$ is much shorter than the characteristic system time $\tau_S$, the Born--Markov approximation leads to the Redfield master equation~\cite{Redfield1957, Breuer2002}. The role of a hierarchy of characteristic times in passing from the microscopic Hamiltonian to an effective Hamiltonian and a white-noise Markov description has been discussed explicitly in quantum-optical settings~\cite{TrubilkoBasharov2020}. With an additional secular approximation, the Redfield equation can be brought to the Gorini--Kossakowski--Sudarshan--Lindblad (GKSL) form~\cite{Lindblad1976, Gorini1976, Breuer2002}; unified weak-coupling GKSL equations can also retain nonsecular terms associated with nearly degenerate Bohr frequencies~\cite{Trushechkin2021}. When the time scales are comparable \cite{deVega2017, Breuer2016}, or when non-Markovian effects must be retained, projection methods are used, including the Nakajima--Zwanzig integrodifferential equation~\cite{Nakajima1958,Zwanzig1960} or time-convolutionless equations \cite{Breuer2002, Shibata1977, Chaturvedi1979}.

	At present, one of the most widely used methods for solving non-Markovian dynamics is the \textit{hierarchical equations of motion} (HEOM) method, introduced by Tanimura and Kubo~\cite{Tanimura1989} and further developed to describe photosynthetic complexes~\cite{Ishizaki2009}. Its main advantage is the ability to obtain numerically converged results that approach the exact system dynamics at arbitrary system--environment coupling strength, provided that the environmental correlation function admits an accurate exponential decomposition and the hierarchy is converged~\cite{Tanimura2020}. The standard equilibrium formulation of HEOM assumes that the environment is initially in a stationary thermal state. The corresponding mode {populations} are given by the Bose--Einstein distribution
	\begin{equation*}
		\braket{N_\omega}_{\mathrm B}=\frac{1}{e^{\beta\omega}-1},
	\end{equation*}
	where $N_\omega$ is the occupation-number operator of a mode at frequency $\omega$ and $\beta$ is the inverse temperature. Here and throughout this work, we set $\hbar=k_{\mathrm B}=1$.
	 Even though standard reduced-system HEOM can account for the influence of system--bath coupling on the spin dynamics, it does not explicitly provide the time evolution of individual bosonic mode occupations.

	An important example is the \textit{spin-boson model}~\cite{Leggett1987, Caldeira1983}, which describes a two-level system interacting with a bosonic environment. It applies to tunneling defects in solids~\cite{Leggett1987, Weiss2012}, molecular dissipative systems~\cite{Nitzan2006}, and superconducting qubits~\cite{Krantz2019, Wendin2017}. Its key characteristic is the spectral density $J(\omega)$. Besides Ohmic spectra, structured environments with a resonant peak have long been studied through damped-mode and reaction-coordinate mappings~\cite{Garraway1997, Thorwart2004}. When the coupling is transverse and the rotating-wave approximation is applicable, an isolated resonant mode reduces to the Jaynes--Cummings model~\cite{JaynesCummings1963}.

		With the development of nonequilibrium thermodynamics~\cite{Vinjanampathy2016, Alicki2018},
		considerable attention has been paid to the dynamics of environments in open quantum systems.
		The frequency-resolved mean mode occupations $\braket{N_\om}(t)$ may depend on both frequency
		and time~\cite{portugal2022effective, kundu2021time, zhang2010effect}. Such time dependence
		can arise even when the bath is initially thermal, due to its interaction with the system.
		Nonthermal initial mode-occupation profiles provide another case of interest. More general
		Gaussian preparations, such as squeezed states~\cite{link2022non, sukharnikov2026non},
		lie outside the diagonal ansatz used in this work.

	For example, Ref.~\cite{kundu2021time} used real-time path integrals to track selected bath coordinates. Gribben \textit{et al.} reconstructed Gaussian-environment correlations and the frequency-resolved redistribution of heat from exact system correlations~\cite{Gribben2022}. Effective environmental-mode occupations have also been studied using HEOM, pseudomodes, and reaction-coordinate mappings~\cite{Lambert2019}, while different nonequilibrium bath preparations have been analyzed with variational and HEOM methods~\cite{Chen2023}. Existing methods for nonequilibrium environments, including generalized versions of HOPS~\cite{link2022non, sukharnikov2026non} and pseudomode approaches~\cite{menczel2024non}, commonly require either a nontrivial many-body calculation or a tractable representation of the environmental correlation function.
	
	In this work, we apply the method developed in Ref.~\cite{meretukov2024time}, based on the Kawasaki--Gunton projection operator~\cite{Kawasaki1973, Robertson1966, Grabert1982}, to the spin-boson model. We derive a computationally inexpensive weak-coupling closure for  the environmental occupations $\braket{N_\om}(t)$ coupled to the spin expectation values $\braket{\sg_\alpha^{(p)}}$. The bath ansatz is Gaussian, diagonal in the mode basis, and has vanishing first moments; the spin ansatz factorizes between different spins. The resulting equations accept thermal and nonthermal initial occupations and a general spectral profile without first decomposing the bath correlation function into exponentials.
	The central output of the method is the frequency-resolved evolution of the bath-mode occupations. From it, one can obtain the bath-side redistribution $p(\omega)\delta\braket{N_\omega}(t)$ and its energy-weighted counterpart $\omega p(\omega)\delta\braket{N_\omega}(t)$. These quantities can in principle be accessed in programmable trapped-ion simulators with engineered structured baths and mode-resolved phonon thermometry~\cite{Sun2025, So2026}. Superconducting circuits with engineered electromagnetic or phononic environments provide a complementary setting for controlled spectral profiles~\cite{Magazzu2018, Kitzman2023}. For a thermal initial state, we compare the spin dynamics with a fixed-truncation HEOM calculation.
	
	The method is perturbative and requires sufficiently weak coupling. It does not reconstruct the full environmental state: intermode coherences, anomalous covariances, coherent displacements, and system--bath correlations lie outside the retained set. More general Gaussian states require the relevant set to be enlarged to include coherent displacements, anomalous covariances, and intermode covariances. Such an extension not only introduces additional evolution equations but also modifies the kernels in the existing spin and mode-occupation equations, producing a substantially larger coupled system. The approximation should therefore be viewed as coupled dynamics of selected spin and bath observables within a diagonal Gaussian projection closure.

	\section{Spin--boson model and retained bath observables}

	As an example, we consider the spin-boson model, namely a two-level system embedded in a bosonic environment. The system $S$ consists of $n$ spins, while the environment consists of $N_e$ harmonic oscillators.

	Let $\varrho(t)$ denote the exact spin--boson state. To close the dynamics of selected expectation values, we project $\varrho(t)$ onto the following ansatz (see Ref.~\cite{meretukov2024time} for details): the spin state factorizes between spins, while the bosonic environment is Gaussian with a diagonal covariance matrix and vanishing linear moments. Thus,
	\begin{equation}
		\label{eq:DenMatr}
		\rho_{\mathrm{ans}}(\vec{\mathcal E}) = \rho_S\otimes\rho_B,
	\end{equation}
	where
	\begin{align*}
		\rho_S &= \bigotimes_{j=1}^n \frac{1}{2}
		\left(I_2+\vec r^{(j)}\cdot\vec\sigma^{(j)}\right),\\
		\rho_B &= Z_B^{-1}\exp\!\left(\mf a^T K\mf a\right),
		& K^T&=K.
	\end{align*}
	Here, $\vec r^{(j)}$ is the Bloch vector of the $j$th spin, and $Z_B$ is the normalization constant of the environmental state. Within the chosen ansatz, expectation values of operators associated with different spins factorize. Consequently, the multispin extension is a product-spin closure and does not retain bath-induced interspin correlations.

	The exact state obeys the standard equation
	\begin{equation}
		\label{eq:LiuvEq}
		\dfrac{d\varrho}{dt} = \mathcal{L}\varrho,
	\end{equation}
	where $\mathcal{L} = \mathcal{L}_S + \mathcal{L}_E + \la\mathcal{L}_{SE}$, with its action on an arbitrary state $X$ given by
	\begin{align}
		\label{eq:GenEq}
		\mathcal{L}_SX
		&= -i\left[\sum_{i = 1}^n
		\left(-\dfrac{\DE_i}{2}\sg_x^{(i)}
		+\dfrac{h_i}{2}\sg_z^{(i)}\right),X\right],\\
		\mathcal{L}_EX
		&= -i\left[\sum_{k,l}H_{kl}b_k^\dagger b_l,X\right]\nonumber\\
		\mathcal{L}_{SE}X
		&= -i\left[\sum_{j=1}^n \sg_z^{(j)}\otimes
		\sum_{k = 1}^{N_e}\left(g_{kj}b_k+g_{kj}^*b_k^\dagger\right),X\right].\nonumber
	\end{align}

	We consider a simplified model without intrinsic dissipation in the environment, i.e., we set $\GA_{kl}=0$. In addition, the environmental Hamiltonian is assumed to be diagonal, $H_{kl}=\omega_k\delta_{kl}$.

	The projection formula of Ref.~\cite{meretukov2024time} gives a kinetic equation for the relevant expectation values. We denote their vector by $\vec{\mathcal E}(t)$ and the corresponding operators by $\vec P$. In the interaction picture, $\mathcal L(t)$ below denotes the interaction Liouvillian with the explicit factor $\lambda$ removed. The bilinear parentheses in the third term denote contraction over the components of $\vec{\mathcal E}$:
	\begingroup
	\small
	\begin{align}
		\dfrac{d}{dt}\vec{\mathcal E}(t)
		&= \lambda\Tr\!\Bigl[
		\vec P\mathcal L(t)
		\rho_{\mathrm{ans}}\!\bigl(\vec{\mathcal E}(t)\bigr)\Bigr]
		\nonumber\\
		&\quad+\lambda^2\Tr\!\Biggl[
		\vec P\mathcal L(t)\int_{t_0}^t\!dt_1\,
		\mathcal L(t_1)\nonumber\\[-2pt]
		&\hspace{20mm}\times
		\rho_{\mathrm{ans}}\!\bigl(\vec{\mathcal E}(t)\bigr)\Biggr]
		\nonumber\\
		&\quad-\lambda^2\Tr\!\Biggl\{\vec P
		\Biggl(
		\Tr\!\Bigl[\vec P\mathcal L(t)
		\rho_{\mathrm{ans}}(\vec{\mathcal E})\Bigr],
		\frac{\partial\rho_{\mathrm{ans}}(\vec{\mathcal E})}
		{\partial\vec{\mathcal E}}
		\Biggr)\nonumber\\[-2pt]
		&\hspace{13mm}\times
		\Biggl(
		\Tr\!\Biggl[\vec P\int_{t_0}^t\!dt_1\,
		\mathcal L(t_1)\nonumber\\[-2pt]
		&\hspace{25mm}\times
		\rho_{\mathrm{ans}}(\vec{\mathcal E})\Biggr],
		\frac{\partial\rho_{\mathrm{ans}}(\vec{\mathcal E})}
		{\partial\vec{\mathcal E}}
		\Biggr)\Biggr\}_{\vec{\mathcal E}=\vec{\mathcal E}(t)}.
		\label{eq:secOrderEqE}
	\end{align}
	\endgroup

	\section{Projection-operator derivation}
	\subsection{Discrete-mode equations}

	The Liouvillian $\mathcal L$ in Eq.~\eqref{eq:secOrderEqE} is written in the interaction picture, whereas the operators in Eq.~\eqref{eq:GenEq} are given in the Schrödinger picture. To transform between these representations, we use the results of Appendix~\ref{ap:A}.

	The spin operators in the interaction picture are
	\begin{align}
		\label{eq:spinOpInt}
		\sg_z^{(p)}(t) = \vec c_p(t)\cdot\vec\sg^{(p)},
	\end{align}
	where $\vec c_p(t)=(c_{xp}(t),c_{yp}(t),c_{zp}(t))$, and the coefficients $c_{\alpha p}(t)$ are defined in Eq.~\eqref{eq:SpinSol}.

	The bosonic operators have the simple dynamics
	\begin{align}
		\label{eq:bosonOpInt}
		b_n(t) &= e^{-iH_{nn}t}b_n=e^{-i\om_n t}b_n,\nonumber\\
		b_n^\dagger(t) &= e^{iH_{nn}t}b_n^\dagger
		=e^{i\om_n t}b_n^\dagger.
	\end{align}

	Combining Eqs.~\eqref{eq:spinOpInt} and \eqref{eq:bosonOpInt}
	and introducing
	\begin{align*}
		S_p(t)&=\vec c_p(t)\cdot\vec\sigma^{(p)},\\
		E_p(t)&=\sum_n\left(g_{np}e^{-i\omega_nt}b_n
		+g_{np}^*e^{i\omega_nt}b_n^\dagger\right),
	\end{align*}
	we obtain
	\begin{equation}
		\label{eq:genInIntRep}
		\mathcal{L}_{SE} = -i\sum_p[S_p(t)\otimes E_p(t),\,\cdot\,].
	\end{equation}
	
	In practice, applying the operator $\mathcal L(t)$ directly to $\rho$ is inconvenient. We therefore introduce the adjoint operator $\mathcal L^*(t)$, defined by
	
	\begin{equation}
		\label{eq:ConjL}
		\Tr(P\mathcal{L}(t)\rho) = \Tr\{\rho\mathcal{L}^*(t)(P)\}.
	\end{equation} 
	This operator acts on the \textit{relevant observables} $\vec P$ rather than on the density matrix $\rho$, thereby simplifying the expressions in Eq.~\eqref{eq:secOrderEqE}. The operator defined by Eq.~\eqref{eq:ConjL} is
	
	\begin{equation}
		\label{eq:GenInIntRep}
		\mathcal{L}_{SE}^*(t) = i\sum_p[S_p(t)\otimes E_p(t),\,\cdot\,].
	\end{equation}
	
	Using the identity $[A\otimes B,\,C\otimes D]=[A,C]\otimes BD+CA\otimes[B,D]$ in Eq.~\eqref{eq:GenInIntRep}, we obtain
	\begin{align}
		\label{eq:EqForAv}
		\Tr\!\left(P_S\mathcal{L}_{SE}\rho\right)
		&=\Tr\!\left[\rho\,\mathcal{L}_{SE}^*(P_S\otimes I)\right]\nonumber\\
		&=\sum_p\braket{i[S_p,P_S]\otimes E_p},\nonumber\\
		\Tr\!\left(P_E\mathcal{L}_{SE}\rho\right)
		&=\Tr\!\left[\rho\,\mathcal{L}_{SE}^*(I\otimes P_E)\right]\nonumber\\
		&=\sum_p\braket{iS_p\otimes[E_p,P_E]},
	\end{align}
	where $P_S$ and $P_E$ are observables in the spin and bosonic spaces, respectively. Below, we use $P_S=\sigma_\alpha^{(j)}$ and $P_E=N_n=b_n^\dagger b_n$. We denote the mean occupation number of a mode by $\braket{N_n}(t)$. Both right-hand sides of Eqs.~\eqref{eq:EqForAv} are linear in the operators $b_m$ and $b_m^\dagger$, which we write compactly as
	\begin{equation}
		\label{eq:EqForAvProp}
		[S_p,P_S]\otimes E_p=O(b,b^\dagger),
		\qquad
		S_p\otimes[E_p,P_E]=O(b,b^\dagger).
	\end{equation}

	Because the environmental ansatz is an undisplaced Gaussian state with a diagonal covariance matrix, all contributions linear in $b_n$ and $b_n^\dagger$ vanish upon averaging. Consequently, the first and third terms in Eq.~\eqref{eq:secOrderEqE} are zero, and only the second term must be evaluated.
	
	Using Eq.~\eqref{eq:GenInIntRep} and the observations above, we finally obtain the equation for the expectation value of a relevant operator:
	
	\begin{align}
		\label{eq:EqForPComms}
		\dfrac{d}{dt}\braket{P}
		&=-\la^2\int_{t_0}^t\!dt_1\sum_{pp'}
		\Bigl\langle[S_{p'}'\otimes E_{p'}',\nonumber\\
		&\hspace{27mm}[S_p\otimes E_p,\,P]]\Bigr\rangle,
	\end{align}
	where $S'_p=S_p(t_1)$ and $E'_p=E_p(t_1)$.

	The equation for the system density matrix (the \textit{Redfield equation}) is
	\begin{align}
		\label{eq:RedfForRhoS}
		\frac{d}{dt}\rho_S
		&=-\lambda^2\int_{t_0}^t\!dt_1\sum_{p,p'}\Bigl\{
		C_{pp'}(t,t_1)\nonumber\\
		&\quad\times[S_p(t),S_{p'}(t_1)\rho_S(t)]
		-C_{p'p}(t_1,t)\nonumber\\
		&\quad\times[S_p(t),\rho_S(t)S_{p'}(t_1)]\Bigr\}.
	\end{align}

	\subsection{Equation for the mean spin operator}
	Substituting the spin operator $P=\sigma_\alpha^{(p)}\otimes I$ into Eq.~\eqref{eq:EqForPComms}, we obtain
	\begin{align}
		\label{eq:EqForSg}
		\frac{d}{dt}\braket{\sg_\al^{(p)}}
		&=2\la^2 \int_{t_0}^t\!dt_1 \Biggl\{
		\sum_{\be\ga}c_{\be p}(t_1)c_{\ga p}(t)\nonumber\\
		&\quad\times\Bigl[
		\bigl(\delta^{\al\be}\braket{\sg_\ga^{(p)}}
		-\delta^{\be\ga}\braket{\sg_\al^{(p)}}\bigr)
		C_p^{+}(t,t_1)\nonumber\\
		&\qquad+i\ve^{\ga\al\be}C_p^{-}(t,t_1)\Bigr]
		-i\sum_{i\neq p}\sum_{\be\ga\de}
		c_{\be i}(t_1)c_{\ga p}(t)\nonumber\\
		&\quad\times\ve^{\ga\al\de}
		\braket{\sg_\be^{(i)}}\braket{\sg_\de^{(p)}}\nonumber\\
		&\qquad\times\bigl[C_{ip}(t_1,t)-C_{pi}(t,t_1)\bigr]
		\Bigr\}.
	\end{align}
	where $C_{ip}(t,t_1)=\braket{E_i(t)E_p(t_1)}$ and
	$C_p^\pm(t,t_1)=C_{pp}(t,t_1)\pm C_{pp}(t_1,t)$.
	For Hermitian environmental operators, $C_{pi}(t_1,t)=C_{ip}^*(t,t_1)$.
	 Here we use the fact that the spin ansatz is a product of individual-spin density matrices, hence, for $i\ne p$, the expectation value $\langle\sigma_\beta^{(i)}\sigma_\delta^{(p)}\rangle$ in Eq.~\eqref{eq:EqForSgn} factorizes as $\langle\sigma_\beta^{(i)}\rangle\langle\sigma_\delta^{(p)}\rangle$.
	A detailed derivation of this expression is given in Appendix~\ref{ap:B}.

	\subsection{Equation for the {mean mode occupation number}}
	Similarly, substituting the bosonic operator $P=I_2\otimes N_m$ into Eq.~\eqref{eq:EqForPComms}, we obtain
	\begin{align}
		\label{eq:EqForN}
		\frac{d}{dt}\braket{N_m}
		&=2\lambda^2\int_{t_0}^t\!dt_1\Bigg\{
		\sum_i\Bigg[\nonumber\\
		&\quad(2\braket{N_m}+1)
		\Im\!\bigl[g_{mi}^*(t_1)g_{mi}(t)\bigr]\nonumber\\
		&\quad\times\bigl[\vec c_i(t_1)\times\vec c_i(t)\bigr]
		\cdot\braket{\vec\sigma^{(i)}}
		+\vec c_i(t_1)\cdot\vec c_i(t)\nonumber\\
		&\quad\times\Re\!\bigl[g_{mi}(t_1)g_{mi}^*(t)\bigr]
		\Bigg]
		+\sum_{i\neq j}\sum_{\alpha\beta}
		c_{\alpha i}(t_1)c_{\beta j}(t)\nonumber\\
		&\quad\times\braket{\sigma_\alpha^{(i)}}
		\braket{\sigma_\beta^{(j)}}\nonumber\\
		&\qquad\times\Re\!\bigl[g_{mi}(t_1)g_{mj}^*(t)\bigr]
		\Bigg\}.
	\end{align}
	 Here, the same reasoning that we used for the product of spin expectation values in Eq.\eqref{eq:EqForSg} applies.

	Eq.~\eqref{eq:RedfForRhoS}, the equivalent equations for the expectation values~\eqref{eq:EqForSg}, and Eq.~\eqref{eq:EqForN} determine the dynamics of the system and the {mean mode occupation numbers} for specified initial conditions.

	 \section{Frequency-resolved bath occupations and coupled spin equations}
	In the preceding section, we obtained the dynamical equations for an arbitrary number of spins within the product-spin closure of Eq.~\eqref{eq:DenMatr}. To expose the bath-side physics, we henceforth restrict the analysis to a single spin.
	
	The equations for the single-spin system are
	\begin{widetext}
	\begin{equation}
		\label{eq:EqForn1}
	\begin{aligned}
		\dfrac{d}{dt}\braket{\sg_\al}
		&=2\la^2\!\int_{t_0}^t\!dt_1\sum_{lk}c_k(t)c_l(t_1)
		\Biggl[
		\sum_{\be\ga}\ve^{k\al\be}\ve^{l\be\ga}\braket{\sg_\ga}(t)
		\\
		&\quad\times\bigl(C(t,t_1)+C(t_1,t)\bigr)
		+i\ve^{k\al l}\bigl(C(t,t_1)-C(t_1,t)\bigr)\Biggr],\\[3pt]
		\dfrac{d}{dt}\braket{N_m}
		&=2\la^2\int_{t_0}^t\!dt_1\Biggl[
		\Re\bigl(g_m(t)g_m^*(t_1)\bigr)
		\sum_l c_l(t)c_l(t_1)\\
		&\quad+\Im\bigl(g_m(t)g_m^*(t_1)\bigr)(2\braket{N_m}+1)
		\sum_{kl\ga}c_l(t_1)c_k(t)\ve^{lk\ga}\braket{\sg_\ga}(t)\Biggr].
	\end{aligned}
	\end{equation}
	\end{widetext}
	
	We now replace the discrete bosonic-mode index $m$ by the continuous spectral parameter $\om$. The corresponding mean occupation is $\braket{N_\om}(t)$. In the continuum limit, the interaction coefficient is written as $g(\om,t)=\tilde{g}(\om,t)e^{-i\om t}$, where the exponential factor arises from the transformation to the interaction picture and $\tilde{g}(\om,t)$ describes the coupling amplitude. Then $g_m(t)g_m^*(t_1)\to G(\om;t,t_1)$, where
	$G(\om;t,t_1)\equiv\tilde{g}(\om,t)\tilde{g}^*(\om,t_1)e^{-i\om(t-t_1)}$.

	At the same time, the sum over modes becomes an integral, $\sum_m\to\int d\om\,p(\om)$, where $p(\om)$ is the mode density. We introduce the two-time spectral kernel $\mf{J}(\om;t,t_1)\equiv p(\om)G(\om;t,t_1)$, which satisfies $\mf{J}(\om;t,t_1)=\mf{J}^*(\om;t_1,t)$. Its diagonal element determines the instantaneous spectral density,
	$J(\om,t)\equiv\mf{J}(\om;t,t)=p(\om)|\tilde{g}(\om,t)|^2$.

	In this notation, the environmental correlation functions take the form
	
	\begin{align}
		\label{eq:ContC}
		C(t,t_1)
		&=\int d\om\,\Bigl[(2\braket{N_\om}(t)+1)
		\Re \mf{J}(\om;t,t_1)\nonumber\\
		&\hspace{25mm}+i\Im \mf{J}(\om;t,t_1)\Bigr],\nonumber\\
		C(t_1,t)
		&=\int d\om\,\Bigl[(2\braket{N_\om}(t)+1)
		\Re \mf{J}(\om;t,t_1)\nonumber\\
		&\hspace{25mm}-i\Im \mf{J}(\om;t,t_1)\Bigr].
	\end{align}
	Because the ansatz on the right-hand side of Eq.~\eqref{eq:secOrderEqE} is evaluated at time $t$, the same $\braket{N_\om}(t)$ enters both expressions. Eq.~\eqref{eq:ContC} implies that $C(t_1,t)=C^*(t,t_1)$ and
	\begin{align*}
		C(t,t_1)+C(t_1,t)
		&=2\int d\om\,(2\braket{N_\om}(t)+1)\nonumber\\
		&\quad\times\Re \mf{J}(\om;t,t_1),\\
		C(t,t_1)-C(t_1,t)
		&=2i\int d\om\,\Im \mf{J}(\om;t,t_1).
	\end{align*}
	
	Substituting these relations into Eqs.~\eqref{eq:EqForn1}, we obtain
	\begin{align}
		\dfrac{d}{dt}\braket{\sg_\al}
		&=4\la^2\int_{t_0}^t\!dt_1\int d\om\,
		\sum_{lk}c_k(t)c_l(t_1)\nonumber\\
		&\quad\times\Biggl[
		\sum_{\be\ga}\ve^{k\al\be}\ve^{l\be\ga}
		\braket{\sg_\ga}(t)\nonumber\\
		&\qquad\times(2\braket{N_\om}(t)+1)
		\Re \mf{J}(\om;t,t_1)\nonumber\\
		&\qquad-\ve^{k\al l}\Im \mf{J}(\om;t,t_1)\Biggr].
		\label{eq:EqForSgCont}
	\end{align}
	For the {mean occupation number} of an individual mode, the factor $p(\om)$ does not arise; therefore,
	\begin{align}
		\dfrac{d}{dt}\braket{N_\om}(t)
		&=2\la^2\int_{t_0}^t\!dt_1\,\Biggl[
		\Re G(\om;t,t_1)
		\sum_l c_l(t)c_l(t_1)\nonumber\\
		&\quad+\Im G(\om;t,t_1)(2\braket{N_\om}(t)+1)\nonumber\\
		&\qquad\times
		\sum_{kl\ga}c_l(t_1)c_k(t)\ve^{lk\ga}\braket{\sg_\ga}(t)\Biggr].
		\label{eq:EqForNCont}
	\end{align}

	Let the real coupling amplitude have the form $\tilde{g}(\om,t)=f(t)g_0(\om)$. For real $f(t)$, this gives $J(\om,t)=f^2(t)J_0(\om)$, where $J_0(\om)=p(\om)|g_0(\om)|^2$. We further assume that it varies slowly on the scale of the environmental correlation time $\tau_B$. For the values of $t_1$ that provide the dominant contribution to the memory integral, one may then set $f(t_1)\simeq f(t)$. This adiabatic approximation is valid under the condition
	$\tau_B|\dot f(t)/f(t)|\ll1$ and yields
	\begin{align*}
		\Re G(\om;t,t_1)
		&\simeq |\tilde{g}(\om,t)|^2\cos\om(t-t_1),\\
		\Im G(\om;t,t_1)
		&\simeq-|\tilde{g}(\om,t)|^2\sin\om(t-t_1),\\
		\Re\mf{J}(\om;t,t_1)
		&\simeq J(\om,t)\cos\om(t-t_1),\\
		\Im\mf{J}(\om;t,t_1)
		&\simeq-J(\om,t)\sin\om(t-t_1).
	\end{align*}
	Since $|\tilde{g}(\om,t)|^2=J(\om,t)/p(\om)$, a specified mode density $p(\om)$ leads to the following closed set of equations. The equation for $\braket{N_\om}(t)$ is considered on the support of the mode density, where $p(\om)>0$:
	\begin{widetext}
	\begin{subequations}
	\label{eq:MainEq}
	\begin{align}
		\frac{d}{dt}\langle\sigma_\alpha\rangle
		&=4\lambda^2\int_{t_0}^{t}\!dt_1\int d\omega\,
		J(\omega,t)\sum_{lk}c_k(t)c_l(t_1)\nonumber\\
		&\quad\times\Biggl[
		\sum_{\beta\gamma}\varepsilon^{k\alpha\beta}
		\varepsilon^{l\beta\gamma}\langle\sigma_\gamma\rangle(t)
		\nonumber\\
		&\qquad\times(2\braket{N_\omega}(t)+1)\cos\omega(t-t_1)
		+\varepsilon^{k\alpha l}\sin\omega(t-t_1)\Biggr],
		\label{eq:MainEq:sg}\\[4pt]
		\frac{d}{dt}\braket{N_\omega}(t)
		&=2\lambda^2\int_{t_0}^{t}\!dt_1\,
		\frac{J(\omega,t)}{p(\omega)}\Biggl[
		\cos\omega(t-t_1)\sum_l c_l(t)c_l(t_1)\nonumber\\
		&\quad-\sin\omega(t-t_1)(2\braket{N_\omega}(t)+1)
		\sum_{kl\gamma}c_l(t_1)c_k(t)\nonumber\\
		&\qquad\times\varepsilon^{lk\gamma}
		\langle\sigma_\gamma\rangle(t)\Biggr].
		\label{eq:MainEq:N}
	\end{align}
	\end{subequations}
	\end{widetext}

	\subsection{Physical interpretation and perturbative order counting}

	The component equations~\eqref{eq:MainEq} take a more transparent form in vector notation. Let $\mathbf s(t)=\langle\boldsymbol\sigma\rangle(t)$, $\mathbf c=\mathbf c(t)$, $\mathbf c'=\mathbf c(t_1)$, and $\tau=t-t_1$. Using the standard vector identities for the Levi-Civita tensor, we obtain
	\begin{widetext}
	\begin{subequations}
	\label{eq:MainEqVector}
	\begin{align}
		\dot{\mathbf s}(t)
		&=4\lambda^2\int_{t_0}^{t}\!dt_1\int d\omega\,J(\omega,t)
		\Bigl\{[2\braket{N_\omega}(t)+1]\cos(\omega\tau)\,
		\mathbf c\times[\mathbf c'\times\mathbf s(t)]
		-\sin(\omega\tau)\,\mathbf c\times\mathbf c'\Bigr\},
		\label{eq:MainEqVector:s}\\
		\frac{d}{dt}\braket{N_\omega}(t)
		&=2\lambda^2\frac{J(\omega,t)}{p(\omega)}
		\int_{t_0}^{t}\!dt_1\,
		\Bigl\{\cos(\omega\tau)\,\mathbf c\cdot\mathbf c'
		+[2\braket{N_\omega}(t)+1]\sin(\omega\tau)
		[\mathbf c\times\mathbf c']\cdot\mathbf s(t)\Bigr\}.
		\label{eq:MainEqVector:N}
	\end{align}
	\end{subequations}
	\end{widetext}
	The double-cross-product term is the linear noise kernel acting on the Bloch vector; without a secular approximation it contains dissipative as well as coherent-renormalization contributions. The term involving $\mathbf c\times\mathbf c'$ originates from the bath commutator and represents its response. In a resonant weak-coupling limit, the two terms combine into the usual balance between stimulated absorption, stimulated emission, and spontaneous emission. Eq.~\eqref{eq:MainEqVector:N} also makes the distinction between the collective bath action $J(\omega,t)$ and the coupling of one physical mode, $J(\omega,t)/p(\omega)=|\tilde g(\omega,t)|^2$, explicit.

	A spectral density alone does not uniquely determine the mean occupation assigned to one continuum mode: the decomposition $J(\omega)=p(\omega)|g(\omega)|^2$ must also be specified. We denote the change from the initial occupation by $\delta\braket{N_\omega}(t)\equiv\braket{N_\omega}(t)-\braket{N_\omega}(0)$. At leading order in $\lambda$, multiplication by $p(\omega)$ removes the explicit $1/p(\omega)$ factor from Eq.~\eqref{eq:MainEqVector:N}. The corresponding occupation-density change can be obtained as
	\begin{equation}
		\delta n_B(\omega,t)=p(\omega)\delta\braket{N_\omega}(t)
		\label{eq:BathNumberDensity}
	\end{equation}
	The corresponding energy-density change can then be obtained as
	\begin{equation}
		\delta e_B(\omega,t)=\omega\,\delta n_B(\omega,t).
		\label{eq:BathEnergyDensity}
	\end{equation}
	The mean occupation $\braket{N_\omega}(t)$ itself remains meaningful when $p(\omega)$ represents the actual mode density of a specified physical platform. The leading-order cancellation of $p(\omega)$ need not hold exactly for the self-consistent equations, in which the mode occupations feed back on the spin. In the comparisons between two regimes below, $\Delta\braket{X}(t)\equiv\braket{X}_{\mathrm{modified}}(t)-\braket{X}_{\mathrm{reference}}(t)$ denotes a difference between their trajectories, not a change from the initial state; the reference regime is specified in each figure caption.

	There is also an important perturbative distinction between the bath-side prediction and its feedback on the spin. For a regular weak-coupling expansion at fixed time, Eq.~\eqref{eq:MainEqVector:N} gives $\braket{N_\omega}(t)=\braket{N_\omega}(0)+O(\lambda^2)$. Since $\braket{N_\omega}$ enters Eq.~\eqref{eq:MainEqVector:s} inside an expression already proportional to $\lambda^2$, replacing the evolving occupation by its initial value changes the spin dynamics only at $O(\lambda^4)$. Thus, the leading $O(\lambda^2)$ result is the bath redistribution itself, whereas feeding that redistribution back into the spin equation is a selective self-consistent resummation of higher-order terms. This order counting need not remain uniform on kinetic time scales $t=O(\lambda^{-2})$ or when a resonant mode accumulates an $O(1)$ occupation change.

	In the Markov and secular limits, the same physical content reduces schematically to a frequency-resolved golden-rule balance,
	\begin{equation}
		\begin{aligned}
		\frac{d}{dt}\braket{N_\omega}
		&\propto\lambda^2|g(\omega)|^2
		|\langle e|\sigma_z|g\rangle|^2\\
		&\quad\times\bigl[P_e(\braket{N_\omega}+1)-P_g\braket{N_\omega}\bigr]\\
		&\quad\times\delta(\omega-\Omega),
		\end{aligned}
		\label{eq:GoldenRuleLimit}
	\end{equation}
	where $P_e$ and $P_g$ are the excited- and ground-state populations. Eq.~\eqref{eq:GoldenRuleLimit} is stated to display the resonance and gain--loss structure; the non-Markovian calculations below use the full equations~\eqref{eq:MainEqVector}. It explains why the bath redistribution is concentrated near the transition frequency $\Omega$.

	\section{Numerical method and benchmarks of spin and bath dynamics}\label{sec:numerical-density}
	The frequency-grid parameters in the following paragraph apply to the continuum calculations, not to the finite-mode benchmark below.
	The numerical calculations solve Eqs.~\eqref{eq:MainEqVector} on the interval $\omega\in[0.05,49.8]$. The uniform mode density is $p=N_{\mathrm{modes}}/(\omega_{\max}-\omega_{\min})$, with the effective mode-count normalization $N_{\mathrm{modes}}=200$ for all figures. This physical normalization is independent of the numerical node counts. The mean occupation $\braket{N_\omega}(t)$ is evolved on a fixed population grid (305 or 611 nodes), while the frequency integrals use a separate, denser quadrature grid concentrated according to $J(\omega,0)[2\braket{N_\omega}(0)+1]$ and supplemented with phase-resolution guard nodes. Runs that are subtracted from one another use the same population grid, so occupation differences are evaluated at identical frequencies. Time integration uses the fixed-step TR-BDF2 scheme with relative and absolute fixed-point tolerances $10^{-7}$ and $10^{-9}$, respectively. Negative occupations and Bloch-vector norms above unity are monitored but not clipped. The source code and its complete list of numerical settings are available in Ref.~\cite{Meretukov2026Code}.
	
\begin{figure*}[t!]
		\centering
		\includegraphics[width=0.78\textwidth]{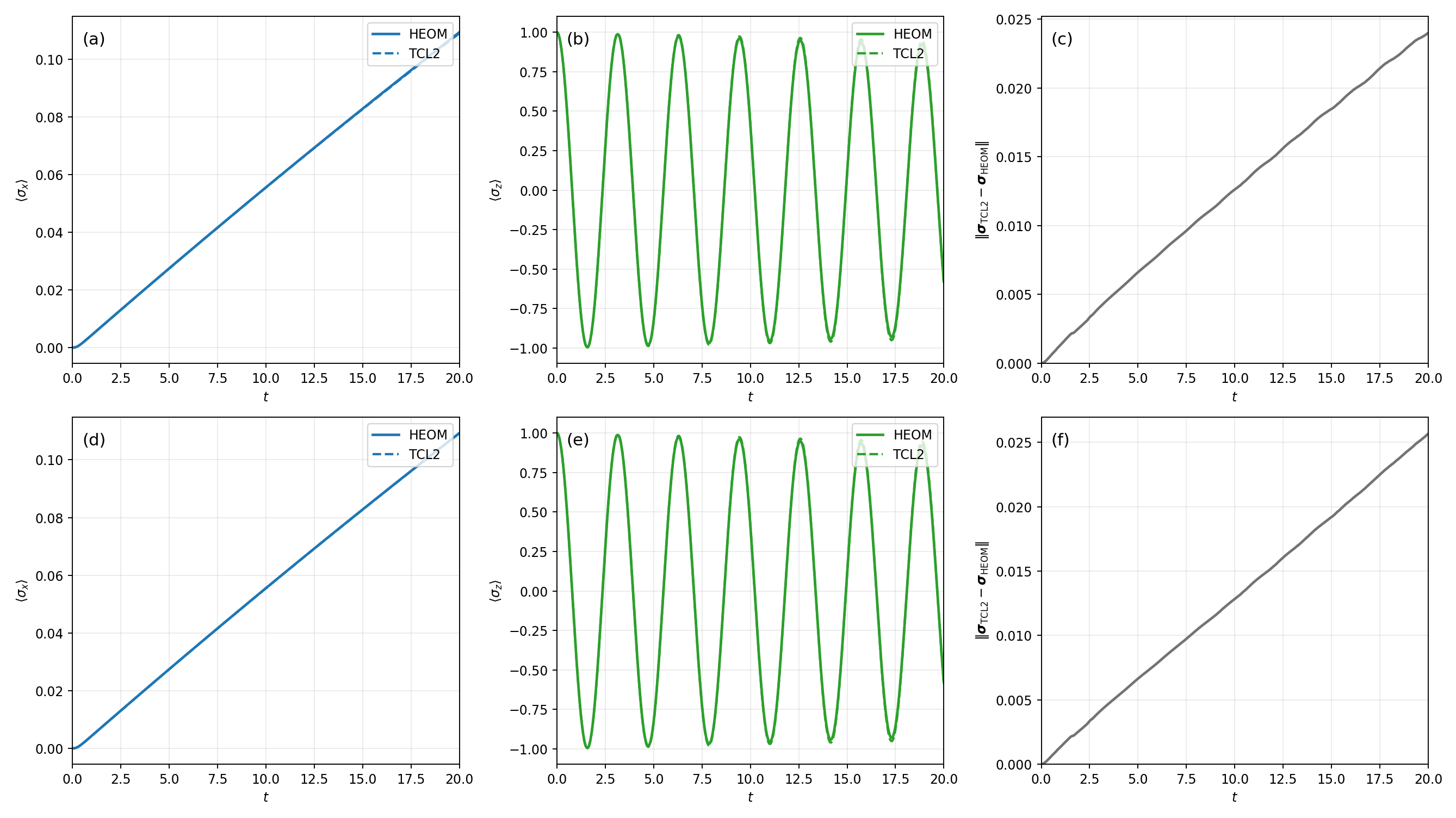}
		\caption{Comparison of HEOM and TCL2 for the stationary Drude--Lorentz density $J_D(\omega)=2\eta\gamma\omega/(\omega^2+\gamma^2)$. Panels (a)--(c) use {dynamically evolving mode populations} in TCL2, whereas panels (d)--(f) use {fixed mode populations}, $\braket{N_\omega}(t)=\braket{N_\omega}(0)$. Panels (a) and (d) show $\langle\sigma_x\rangle$, panels (b) and (e) show $\langle\sigma_z\rangle$, and panels (c) and (f) show the Bloch-vector difference. The parameters are $\Delta=2.0$, $h=0$, $\lambda=0.1$, $\beta=5.0$, $\eta=0.1875$, $\gamma=7.5$, and $0\leq t\leq20$. The initial spin state has $\langle\boldsymbol{\sigma}\rangle(0)=(0,0,1)$, and the bath is initially thermal. The bath parameters are chosen in a regime similar to that considered in Ref.~\cite{kundu2021time}. TCL2 uses 801 time points, a physical mode-count normalization of 200, 305 fixed population-grid nodes, and 5200 adaptive quadrature nodes; HEOM uses hierarchy depth 6 and 20 Pad\'e terms. Here and below $\hbar=k_B=1$.}
		\label{fig:HEOMComparison}
	\end{figure*}

	In Fig.~\ref{fig:HEOMComparison}, the numerical TCL2 solution is compared with a fixed-truncation HEOM reference~\cite{Tanimura2020} for the Drude--Lorentz spectral density $J_D(\omega)=2\eta\gamma\omega/(\omega^2+\gamma^2)$. The HEOM calculation uses the solver provided by QuTiP~5.2.3~\cite{Lambert2023}, hierarchy depth 6, and 20 Pad\'e terms. This comparison is a benchmark against the stated HEOM truncation rather than a claim of numerical convergence. As a limited Pad\'e-order check, a calculation with hierarchy depth 6 and 18 Pad\'e terms differs from the displayed HEOM result by at most 0.0033 in Bloch-vector norm over the same interval. With dynamically evolving mode populations, the Bloch-vector difference from HEOM grows over the plotted interval and is most pronounced in $\braket{\sg_z}$ among the components shown. The unplotted $\braket{\sg_y}$ has a comparable maximum component difference. The dynamic- and frozen-occupation TCL2 curves remain close, with a maximum Bloch-vector difference of 0.0017 between them. For the selected parameters, evolving the mode populations gives a slightly smaller maximum Bloch-vector difference from HEOM than keeping them fixed: 0.0240 compared with 0.0257. Their small difference in spin observables is consistent with fixed-time order counting: the feedback of $\delta\braket{N_\omega}=O(\lambda^2)$ on the spin first appears at $O(\lambda^4)$. The remaining differences from HEOM can arise from the second-order truncation, the diagonal Gaussian closure, and the finite HEOM and frequency cutoffs. For these runs, the TCL2 equation-solver times were 17.2 s with evolving populations and 10.3 s with fixed populations, compared with 4243.4 s for the HEOM reference run. These timings characterize the present implementation and parameter set.  The calculations were performed on a computer equipped with an Intel Core i5-12450H processor and 16 GB of RAM. The reported wall-clock times refer to individual solver runs on this machine.

	\subsection{Finite-mode comparison with the full Hamiltonian}

\begin{figure*}[t!]
		\centering
		\includegraphics[width=0.95\textwidth]{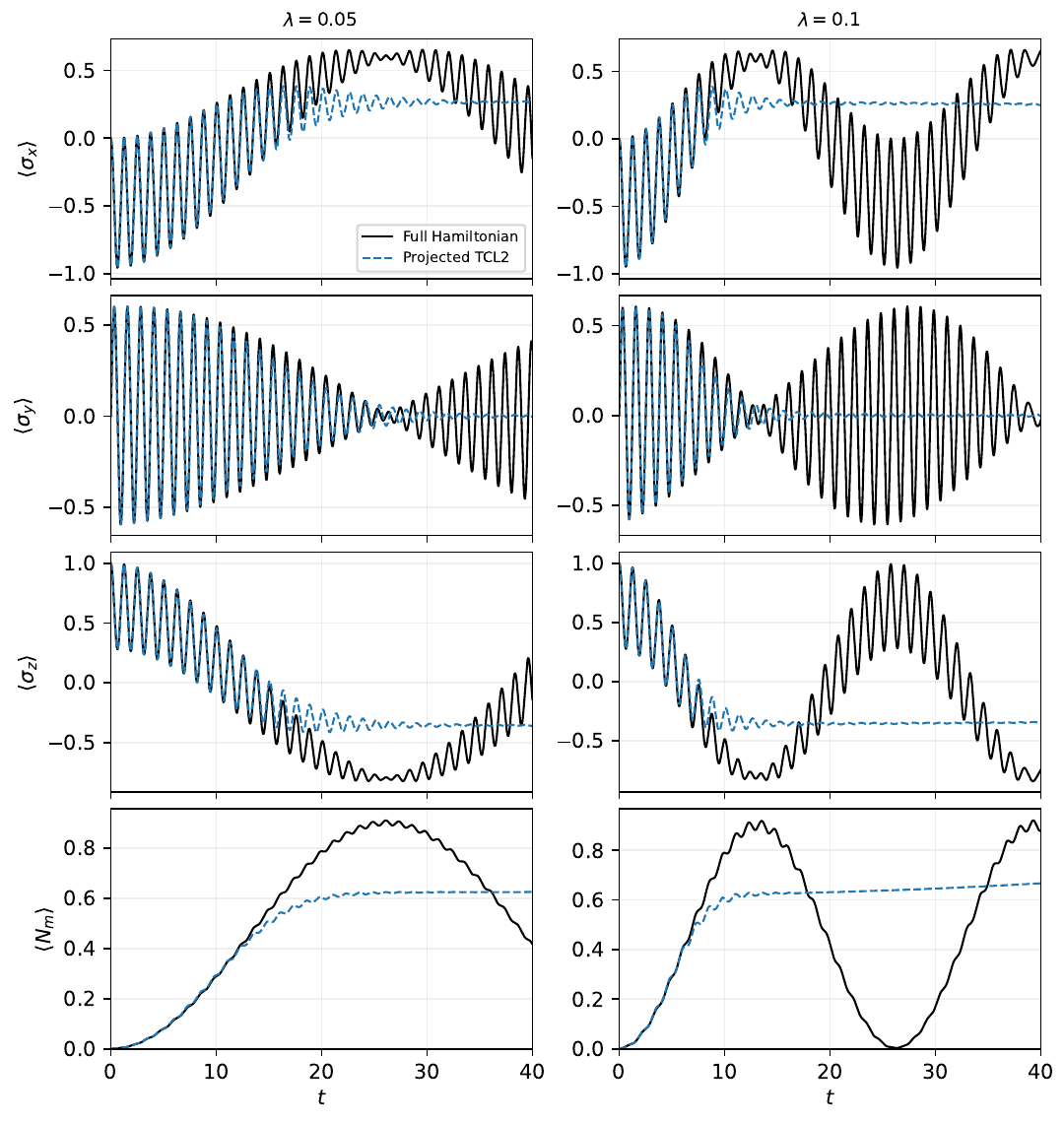}
		\caption{One-mode spin--boson dynamics at resonance, $\omega_m=\Omega=\sqrt{\Delta^2+h^2}=5$. Black solid curves show unitary evolution under the full Hamiltonian (QuTiP); blue dashed curves show the discrete projected equations~\eqref{eq:EqForn1}. The left and right columns correspond to $\lambda=0.05$ and $0.1$, respectively. Rows show the three spin components and the mean mode population $\braket{N_m}$. The common parameters are $\Delta=3$, $h=4$, $g_m=2$, $\beta=5$, and $0\leq t\leq40$, with 4600 output times. The initial spin Bloch vector is $(0,0,1)$, the mode is initially thermal, and the full-model calculation retains 80 Fock states.}
		\label{fig:SingleModeResonant}
	\end{figure*}

\begin{figure*}[t!]
		\centering
		\includegraphics[width=0.95\textwidth]{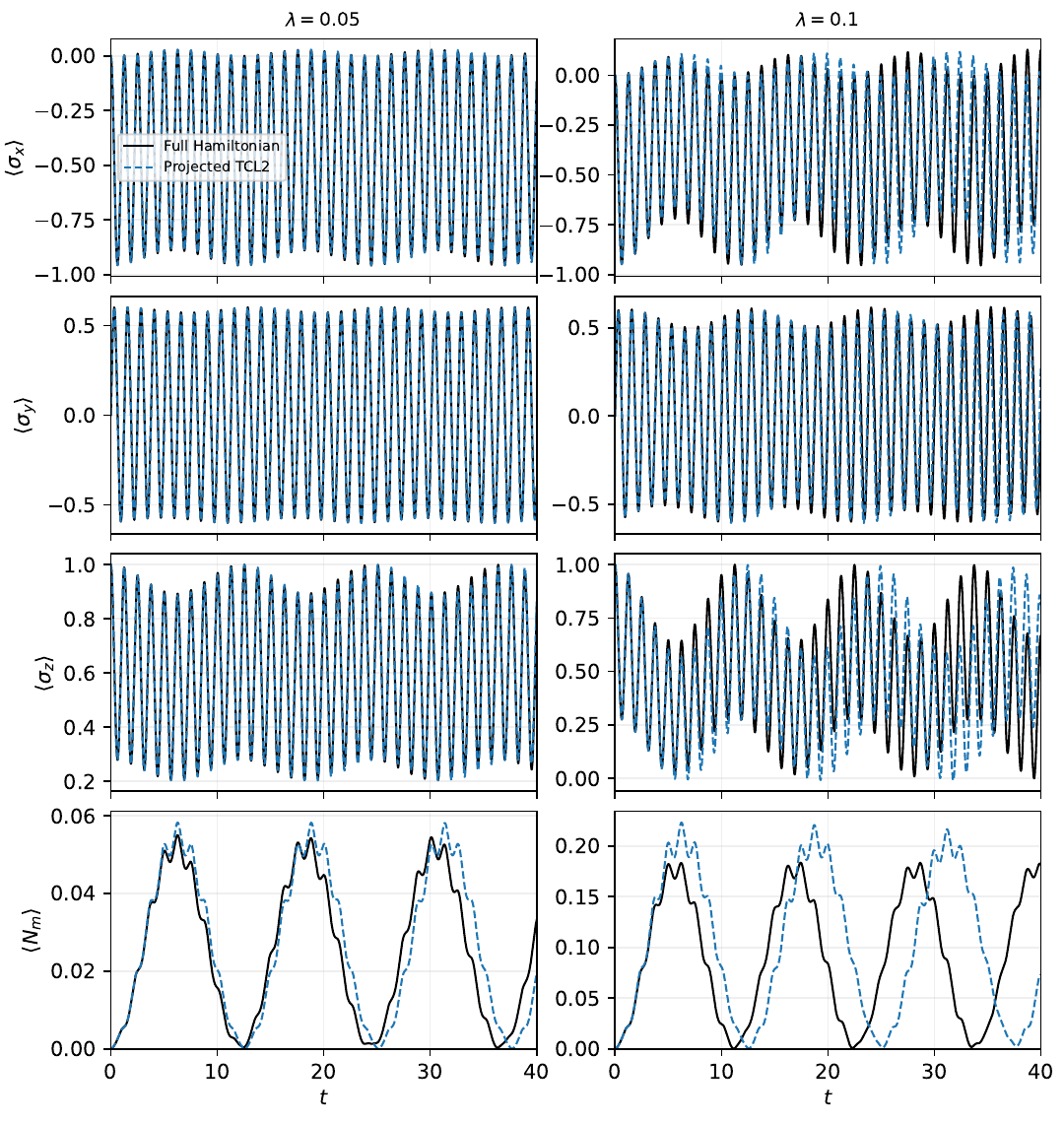}
		\caption{The same one-mode comparison as in Fig.~\ref{fig:SingleModeResonant}, but with $\omega_m=4.5$ while the spin transition frequency remains $\Omega=5$. Left and right columns again show $\lambda=0.05$ and $0.1$; all other physical parameters and initial conditions are unchanged. The full-model calculation retains 60 Fock states, and each run uses 4600 output times over $0\leq t\leq40$.}
		\label{fig:SingleModeDetuned}
	\end{figure*}

	The discrete equations~\eqref{eq:EqForn1} can be compared directly with unitary evolution under the full spin--boson Hamiltonian, without taking a continuum limit. Figures~\ref{fig:SingleModeResonant} and \ref{fig:SingleModeDetuned} show this comparison for one bath mode at resonance, $\omega_m=\Omega=5$, and with detuning, $\omega_m=4.5$, respectively. Each frequency is calculated at both $\lambda=0.05$ and $\lambda=0.1$, with all other physical parameters unchanged. For $\lambda=0.05$, the maximum Bloch-vector difference over $0\leq t\leq40$ decreases from 0.595 at resonance to 0.038 with detuning. At $\lambda=0.1$, the corresponding differences are 1.37 and 0.44. The same comparison for the mode population gives maximum absolute differences of 0.285 and 0.015 at $\lambda=0.05$, and 0.64 and 0.17 at $\lambda=0.1$. Thus, detuning improves agreement for these parameters, but increasing the coupling still produces a substantial difference even off resonance.

	At short times, the projected and full-model trajectories are close. At resonance, the full one-mode model subsequently exhibits pronounced coherent returns that the projected solution does not reproduce. This is consistent with a finite-time limitation of the second-order expansion and the diagonal Gaussian closure; the present comparison does not isolate their separate contributions. The importance of assessing the validity of second-order approximations in spin--boson dynamics has also been discussed in Ref.~\cite{Fruchtman2016}, although that work does not address the present single undamped resonant mode. The populations at the upper Fock level remain negligible in all four full-model runs, so the observed differences cannot be attributed to an occupied Fock cutoff.

	\clearpage

	\section{Frequency-resolved bath dynamics: numerical results}

	\subsection{Resonance-localized redistribution in an Ohmic bath}

\begin{figure*}[t!]
		\centering
		\includegraphics[width=0.88\textwidth]{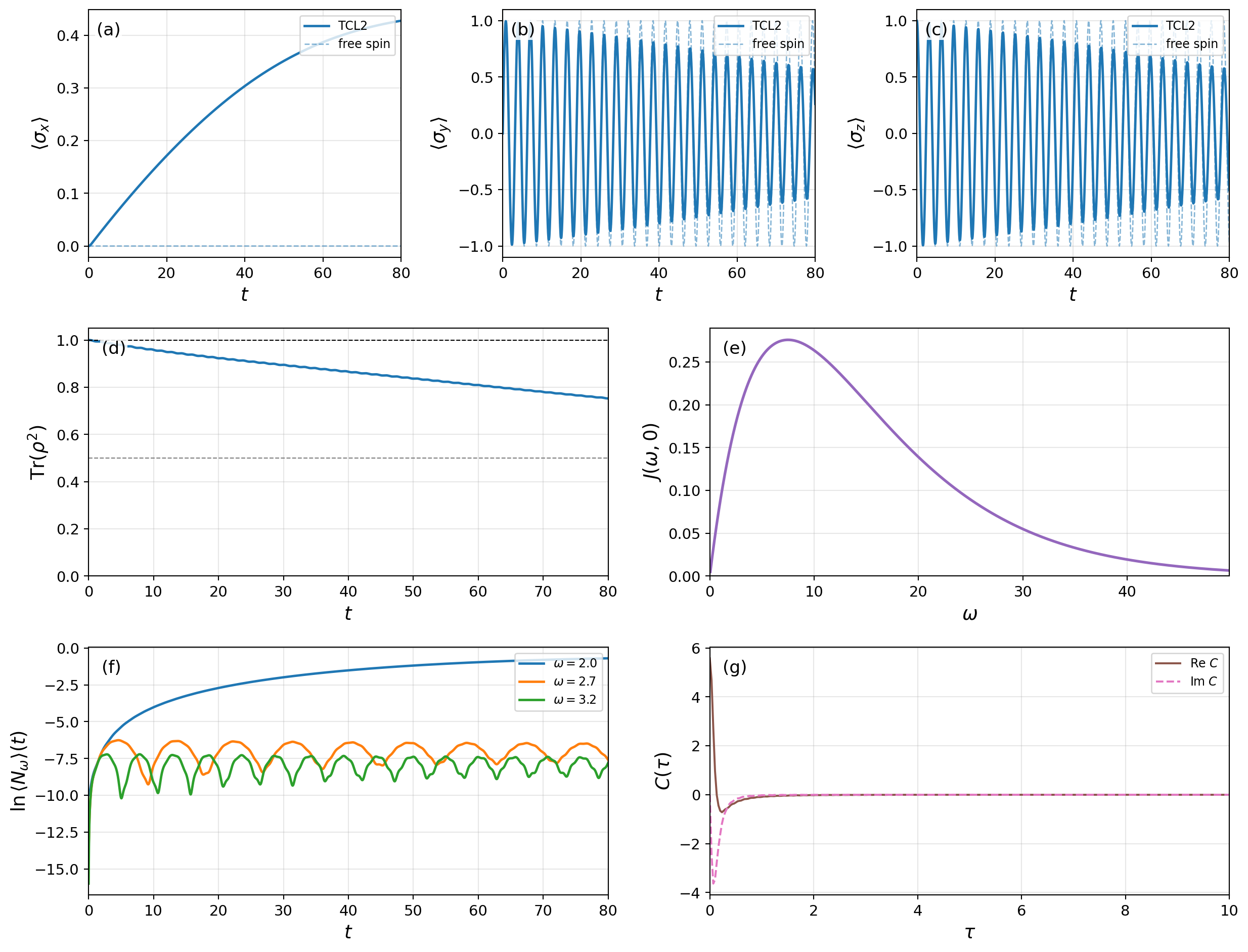}
		\caption{TCL2 dynamics for a stationary Ohmic reservoir on $0\leq t\leq80$ with 3201 time points. The spectral density is $J_O(\omega)=\frac{\xi}{2}\omega\exp(-\omega/\omega_c)$, and the parameters are $\Delta=2$, $h=0$, $\lambda=0.1$, $\xi=0.2$, $\omega_c=7.5$, and $\beta=5$. The initially thermal {population} $\braket{N_\omega}(t)$ is evolved dynamically. The calculation uses a physical mode-count normalization of 200, 611 fixed population-grid nodes, and 5200 adaptive quadrature nodes. The same population grid is used in Figs.~\ref{fig:OhmicLorentz}--\ref{fig:SlowJt}; population differences are evaluated directly at identical frequencies. Panels (a)--(c) show the spin components and free-spin evolution, (d) the purity, and (e) $J_O(\omega)$. Panel (f) shows the selected mode populations at $\omega=2$, $2.7$, and $3.2$ as $\ln\braket{N_\omega}(t)$. Panel (g) shows the initial reservoir correlation function $C(\tau)$.}
		\label{fig:OhmicLowT80}
	\end{figure*}

	Fig.~\ref{fig:OhmicLowT80} presents the dynamics for a time-independent Ohmic spectral density. The occupation change is concentrated near the transition frequency, in agreement with the resonance structure of Eq.~\eqref{eq:GoldenRuleLimit}. The oscillation amplitudes of $\braket{\sg_y}$ and $\braket{\sg_z}$ decay, $\braket{\sg_x}$ increases, and the purity decreases. The mode populations underlying the displayed $\ln\braket{N_\omega}(t)$ correspond to the uniform mode density specified in Sec.~\ref{sec:numerical-density}; the occupation-density redistribution of Eq.~\eqref{eq:BathNumberDensity} can be obtained from the calculated populations by subtracting their initial values and multiplying by $p$.

	\subsection{Enhancement by a structured spectral density}

\begin{figure*}[t!]
		\centering
		\includegraphics[width=0.68\textwidth]{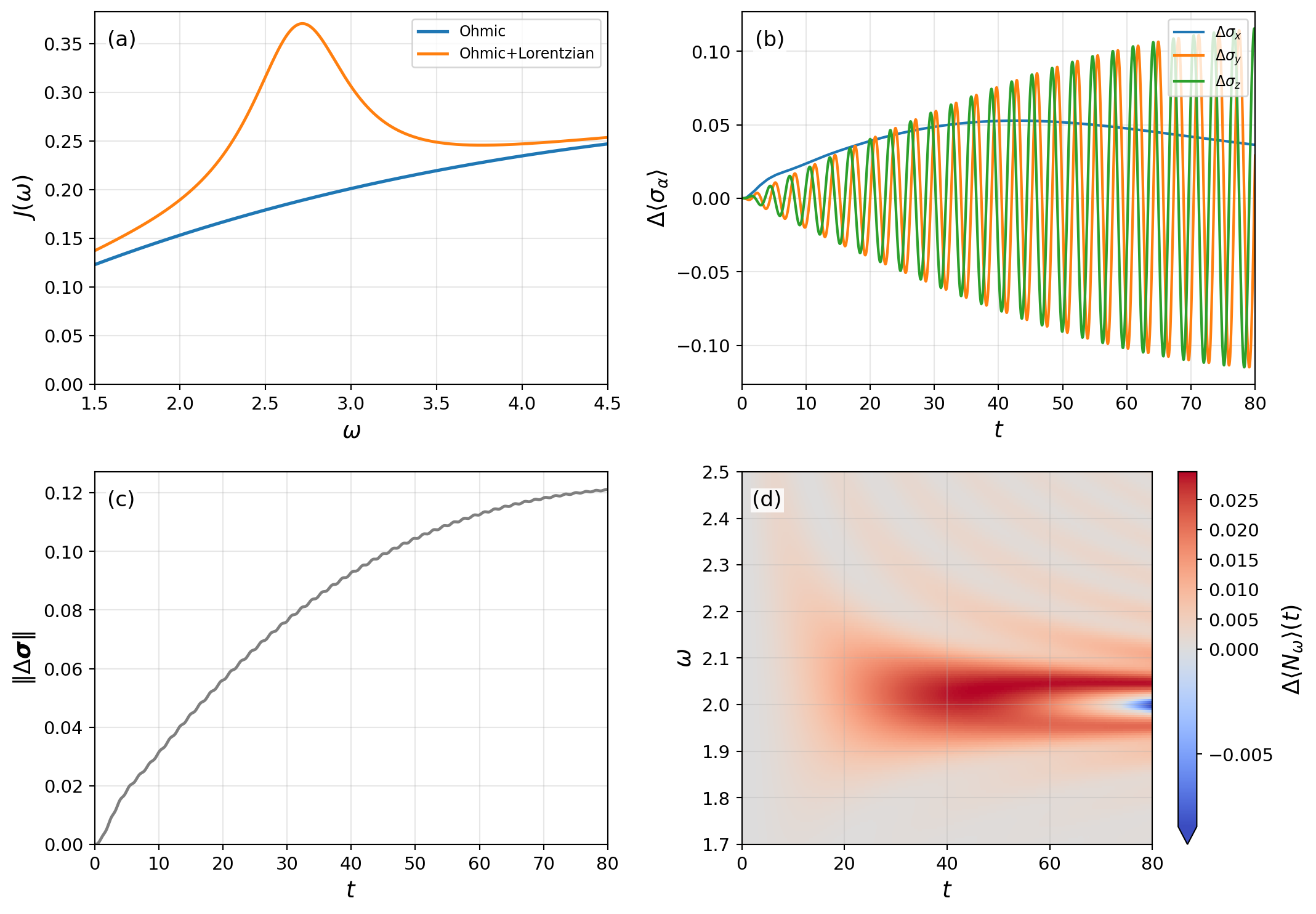}
		\caption{Effect of adding a Lorentzian resonance to the stationary Ohmic density. Panel (a) compares $J_O(\omega)$ with $J_O(\omega)+J_L(\omega)$, where $J_L(\omega)=A_L\Gamma_L/\{\pi[(\omega-\omega_L)^2+\Gamma_L^2]\}$. Panel (b) shows the componentwise change $\Delta\langle\sigma_\alpha\rangle$ relative to the Ohmic calculation in Fig.~\ref{fig:OhmicLowT80}, panel (c) shows $\|\Delta\boldsymbol{\sigma}\|$, and panel (d) shows $\Delta\braket{N_\omega}(t)$ for the uniform physical mode density specified in Sec.~\ref{sec:numerical-density}, relative to the same Ohmic calculation. In panel (d), the lower color limit is set to the first percentile of negative values; more negative values are saturated. The parameters are $\xi=0.2$, $\omega_c=7.5$, $A_L=0.2$, $\omega_L=2.7$, $\Gamma_L=0.35$, $\Delta=2$, $h=0$, $\lambda=0.1$, $\beta=5$, and $0\leq t\leq80$. The initially {thermal mode populations} evolve dynamically.}
		\label{fig:OhmicLorentz}
	\end{figure*}

	Fig.~\ref{fig:OhmicLorentz} demonstrates the change in the dynamics caused by adding a Lorentzian component to the time-independent Ohmic spectral density. The maximum of the Lorentzian component is located near the resonance frequency of the system; therefore, its inclusion noticeably modifies the dynamics compared with the purely Ohmic case. 
	Unlike the Drude--Lorentz spectral density considered in Fig.~\ref{fig:HEOMComparison}, the spectral density used here cannot be treated directly by the particular HEOM implementation employed in this work. Its treatment within HEOM would require an additional decomposition of the bath correlation function into a finite sum of exponentials or an approximate representation in terms of HEOM-compatible spectral components, together with a separate convergence analysis. Thus, treating these regimes within the HEOM implementation used in this work requires additional approximations of the environmental correlation function or spectral density, whereas the proposed method allows the selected spectral density to be directly substituted into the derived equations. Therefore, for this and subsequent regimes, we present the results of the proposed method without a direct comparison with HEOM.
	The difference in the environmental mode populations is predominantly concentrated near the resonance frequency of the system. The differences $\Delta\braket{\sg_y}$ and $\Delta\braket{\sg_z}$ exhibit oscillatory behavior, indicating a modification of the oscillatory spin dynamics that may be associated with changes in both the amplitudes and phases of these components. At the same time, the monotonic growth of $\lVert\Delta\braket{\boldsymbol{\sg}}\rVert$ shows that the overall {difference} between the spin trajectories increases over the considered time interval and is not canceled by the individual components.
	
	\subsection{Response to a nonthermal bath preparation}

\begin{figure*}[t!]
		\centering
		\includegraphics[width=0.68\textwidth]{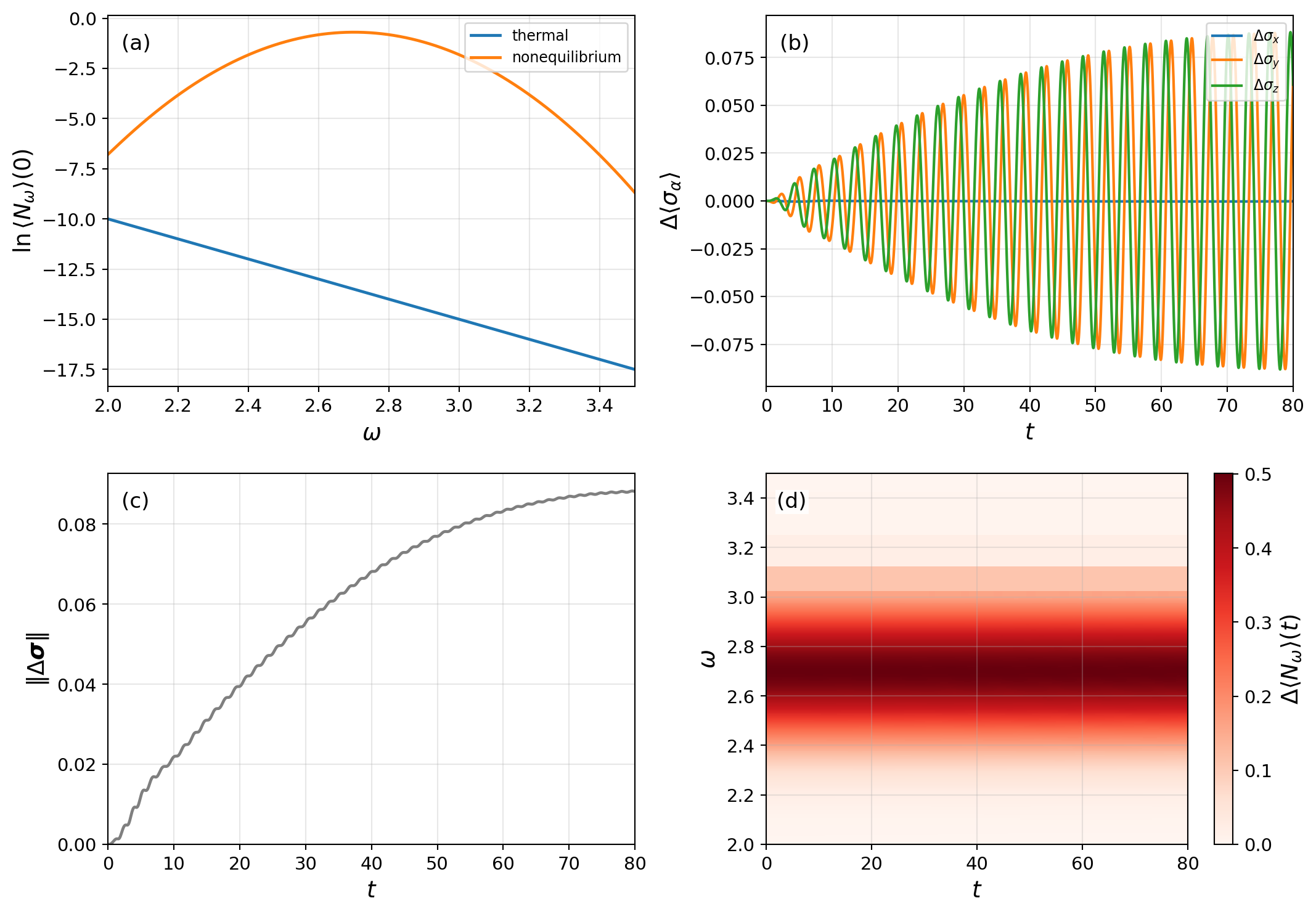}
		\caption{Effect of a nonequilibrium initial mode {population} for the stationary Ohmic-plus-Lorentzian density of Fig.~\ref{fig:OhmicLorentz}. Panel (a) compares $\ln\braket{N_\omega}(0)$ for the thermal and nonequilibrium initial {populations} near the added peak. The latter is $\braket{N_\omega}(0)=[\exp(\beta\omega)-1]^{-1}+A_N\exp[-(\omega-\omega_N)^2/(2s_N^2)]$, with $A_N=0.5$, $\omega_N=2.7$, and $s_N=0.2$. Panels (b)--(d) show, respectively, $\Delta\langle\sigma_\alpha\rangle$, $\|\Delta\boldsymbol{\sigma}\|$, and $\Delta\braket{N_\omega}(t)$ relative to the thermal calculation. The remaining parameters are $\Delta=2$, $h=0$, $\lambda=0.1$, $\beta=5$, and $0\leq t\leq80$.}
		\label{fig:NonequilibriumInitialN}
	\end{figure*}

	Fig.~\ref{fig:NonequilibriumInitialN} presents the numerical results for a system with a nonequilibrium initial state of the environment. The initial mode {populations} are defined as
	\begin{equation*}
	\braket{N_\om}(0)=\braket{N_\om}_{\mathrm B}+N_{\mathrm{add}}(\om),
	\end{equation*}
	where $\braket{N_\om}_{\mathrm B}$ is the Bose--Einstein distribution and $N_{\mathrm{add}}(\om)$ is an additional Gaussian distribution localized near the frequency $\om_N$. The spectral density remains unchanged. The added peak remains clearly distinguishable in the difference between the environmental mode {populations}, $\Delta\braket{N_\om}(t)$, near $\om_N$. The nonequilibrium initial distribution also modifies the dynamics of $\braket{\sg_y}$ and $\braket{\sg_z}$, whereas the dynamics of $\braket{\sg_x}$ remains almost unchanged. This example demonstrates that, within the {chosen} ansatz, the proposed method provides a unified treatment of both equilibrium and nonequilibrium diagonal Gaussian initial states of the environment.
	
	\subsection{Delayed response under slow coupling modulation}

\begin{figure*}[t!]
		\centering
		\includegraphics[width=0.68\textwidth]{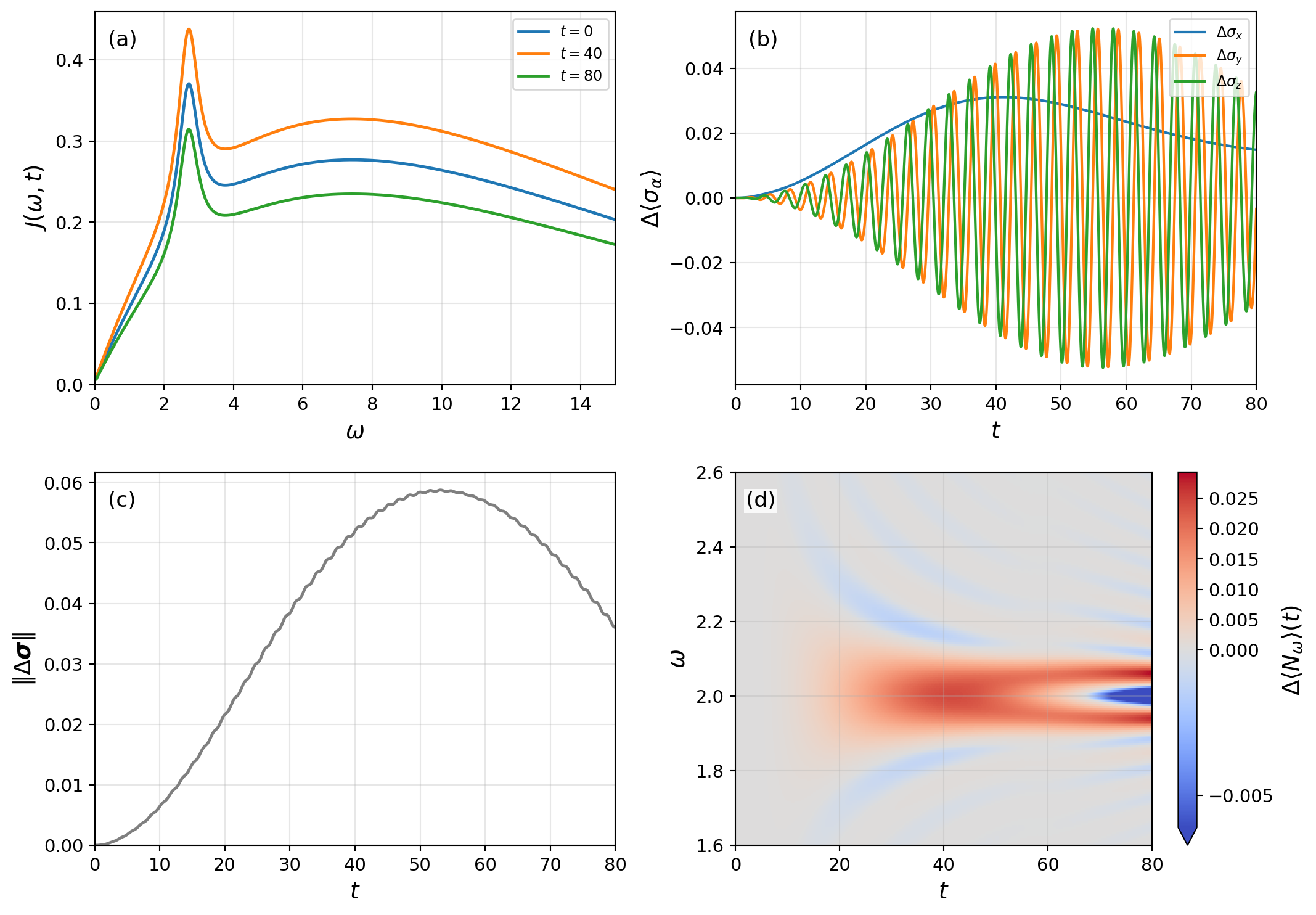}
		\caption{Effect of a slowly modulated spectral density $J(\omega,t)=J_0(\omega)[1+A_t\cos(\nu t+\phi)]$, where $J_0(\omega)$ is the Ohmic-plus-Lorentzian density of Fig.~\ref{fig:OhmicLorentz}. The corresponding coupling-amplitude modulation is $f(t)=\sqrt{1+A_t\cos(\nu t+\phi)}$. Panel (a) shows the density at $t=0$, 40, and 80. Panels (b)--(d) show, respectively, $\Delta\langle\sigma_\alpha\rangle$, $\|\Delta\boldsymbol{\sigma}\|$, and $\Delta\braket{N_\omega}(t)$ relative to the stationary calculation. In panel (d), the lower color limit is set to the first percentile of negative values; more negative values are saturated. The parameters are $A_t=0.2$, $\nu=0.05$, $\phi=-\pi/2$, $\Delta=2$, $h=0$, $\lambda=0.1$, $\beta=5$, and $0\leq t\leq80$. {The mode populations} start from the Bose distribution and evolve dynamically.}
		\label{fig:SlowJt}
	\end{figure*}

	Fig.~\ref{fig:SlowJt} shows the dynamics of the system for a slowly varying spectral density $J(\om,t)=f^2(t)J_0(\om)$. The frequency profile of the spectral density is preserved, while its overall amplitude slowly varies with time in accordance with the adiabatic approximation employed here. Near the resonance frequency, the difference in the mode populations, $\Delta\braket{N_\om}(t)$, initially increases, subsequently decreases, and eventually changes sign relative to the stationary solution. In contrast to the previous regimes, the deviation $\lVert\Delta\braket{\boldsymbol{\sg}}\rVert$ from the solution with a stationary spectral density initially increases and then begins to decrease, with a certain delay, after the amplitude of the spectral density passes through its maximum. This behavior is consistent with a finite response time and non-Markovian memory of the system~\cite{deVega2017, Breuer2016}.

	\section{Mode-resolved observables and energy-balance considerations}

	The closest direct platform is a trapped-ion spin--boson simulator, where internal states represent the spin and collective motional modes represent the bosonic environment. Structured spectra and controlled initial temperatures have already been demonstrated~\cite{Sun2025}. Mode-resolved blue-sideband spectroscopy can reconstruct phonon-number distributions and has been used to follow phonon populations in engineered thermal reservoirs~\cite{So2026}. A direct test of Eqs.~\eqref{eq:MainEqVector} would prepare a selected set of motional modes, switch off the preparation channels during the coherent spin--boson evolution, and measure both spin tomography and the final mode occupations as functions of interaction time. If heating and cooling remain active, the present equations must first be extended to nonzero $\Gamma_{kl}$.

	Engineered electromagnetic and surface-acoustic-wave environments of superconducting qubits provide a complementary realization of Ohmic and structured spectra~\cite{Magazzu2018, Kitzman2023}. In these devices the qubit observables and spectral profile are directly accessible, although a mode-resolved measurement of $\braket{N_\omega}(t)$ is generally less direct than in trapped-ion systems.

	A further physical diagnostic is energy balance. For a stationary Hamiltonian the conserved quantity is $E_{\mathrm{tot}}=E_S+E_B+E_{\mathrm{int}}$, with
	\begin{equation}
		E_B(t)=\int d\omega\,\omega p(\omega)\braket{N_\omega}(t).
		\label{eq:BathEnergy}
	\end{equation}
	The interaction energy $E_{\mathrm{int}}=\lambda\langle V\rangle$ contains system--bath correlations not retained in the present ansatz and can contribute at $O(\lambda^2)$. Consequently, conservation of $E_S+E_B$ alone is not expected at that order. A quantitative energy-balance test requires either a perturbative reconstruction of $\langle V\rangle$ or an enlarged relevant-observable set. For a time-dependent coupling, the external power $\langle\partial_tH\rangle$ must additionally be separated from the bath-energy change.

	\section{Conclusion}
	
	We derived weak-coupling equations for frequency-resolved bosonic occupations  and for spin observables within a product-spin, diagonal-Gaussian projection closure. The equations apply to an arbitrary number of spins within this factorization, while the numerical examples concern one spin. General stationary spectral profiles enter directly, and a slowly varying coupling is included through $J(\omega,t)=f^2(t)J_0(\omega)$.
	
	From the calculated frequency-resolved occupations, changes in the occupation and energy densities $p(\omega)\delta\braket{N_\omega}(t)$ and $\omega p(\omega)\delta\braket{N_\omega}(t)$ can also be obtained; the occupation of one mode additionally depends on the chosen decomposition $J=p|g|^2$. The numerical results also show a resonant redistribution concentrated near the spin transition frequency. A Lorentzian spectral feature enhances the response in its resonance region, while a nonthermal initial occupation changes the spin trajectory without changing $J$. Slow modulation produces a delayed bath and spin response within the adiabatic approximation.
	
	For the Drude--Lorentz spectrum, both TCL2 variants reproduce the main oscillatory features of the spin dynamics, but the dynamic-occupation variant is slightly closer to the fixed-truncation HEOM reference for the selected parameters. Their small difference is consistent with perturbative order counting: at fixed time, the feedback of $\delta\braket{N_\omega}=O(\lambda^2)$ on the spin first enters at $O(\lambda^4)$. The self-consistent feedback retained by the coupled equations should therefore be regarded as a selective higher-order resummation rather than a uniformly controlled second-order effect.

	In the single-mode comparisons shown in Figs.~\ref{fig:SingleModeResonant} and \ref{fig:SingleModeDetuned}, the TCL2 solution agrees well with unitary evolution at short times both at resonance and for a small detuning. At longer times, the detuned solution retains the overall oscillatory character of the unitary dynamics despite quantitative differences, whereas the resonant projected solution does not reproduce the coherent recurrences of the full model. Because the projected calculation also uses a diagonal Gaussian ansatz for the bath, this comparison does not isolate the effect of the second-order truncation. It illustrates a limit of the combined approximation at long times rather than an error in the derived equations.
	
	The principal advantage of the method is inexpensive access to bath-side observables for thermal and nonthermal diagonal Gaussian preparations without an exponential decomposition of the correlation function. Its main limitations are weak coupling, the restricted Gaussian closure, the absence of explicit system--bath and interspin correlations, and the adiabatic treatment of time-dependent couplings.
	
	Further development should include controlled higher-order corrections, anomalous and off-diagonal bath covariances, reconstruction of interaction energy, and multispin correlations. These extensions would also permit direct comparison with driven and dissipative bosonic modes in current quantum-simulation platforms.
		
	\begin{acknowledgments}
	This work was supported by Russian Science Foundation Grant No.~25-21-00700, \url{https://rscf.ru/project/25-21-00700/}.
	\end{acknowledgments}

	\section*{Data and Code Availability}

	The source code and configuration format used for the numerical calculations are available in Ref.~\cite{Meretukov2026Code}. The numerical data underlying the figures are available from the authors upon reasonable request.

	\clearpage

\appendix
\setcounter{equation}{0}
\renewcommand{\theequation}{\Alph{section}\arabic{equation}}

	\section{Transformation to the interaction picture}\label{ap:A}

	To transform to the interaction picture, one must solve the following evolution equation for $A$:
	\begin{equation*}
		\dfrac{\pr}{\pr t}A(t) = -\mathcal{L}_0A(t),
	\end{equation*}
	where $\mathcal L_0$ is the Liouvillian of the free dynamics,
	\begin{align}
		\label{eq:FreeDynGen}
		\mathcal{L}_0A
		&=-i\Biggl[\sum_{i = 1}^n
		\left(-\dfrac{\DE_i}{2}\sg_x^{(i)}
		+\dfrac{h_i}{2}\sg_z^{(i)}\right)\nonumber\\
		&\hspace{18mm}+\sum_{k,l}H_{kl}b_k^\dagger b_l,A\Biggr].
	\end{align}
	
	Because the free Hamiltonian is a sum of independent single-spin Hamiltonians, the transformation of each spin operator is independent of the other spins. Furthermore, Eq.~\eqref{eq:FreeDynGen} contains no mixed spin-boson terms, so the free dynamics of the system and the environment can be considered separately.
	\subsection{Spin operators}
	As noted above, it is sufficient to consider a single spin, i.e., to set $n=1$, in order to determine the transformation of an individual spin operator.

	Taking into account the independence of the free system and environmental dynamics, the equation for $\sg_\al(t)$ is
	\begin{equation*}
		\dfrac{\pr}{\pr t}\sg_\al = \left[i\!\left(-\dfrac{\DE}{2}\sg_x+\dfrac{h}{2}\sg_z\right),\sg_\al\right]
	\end{equation*}
	Applying the commutation relations $[\sg_j,\sg_k]=2i\,\ve_{jkm}\sg_m$, we obtain
	\begin{align}
		\label{eq:IntRepEq}
		\dfrac{\pr}{\pr t}\sg_x&=-h\sg_y,\nonumber\\
		\dfrac{\pr}{\pr t}\sg_z&=-\DE\sg_y,\nonumber\\
		\dfrac{\pr}{\pr t}\sg_y&=\DE\sg_z+h\sg_x
	\end{align}
	
	Solving Eqs.~\eqref{eq:IntRepEq}, we obtain the following interaction-picture operators:
	
	\begin{widetext}
	\begin{align}
		\label{eq:SpinSol}
		\sg_x^{(i)}(t) &= \sg_x^{(i)}\frac{\DE_i^2+h_i^2\cos(t\OM_i)}{\OM_i^2}
		- \sg_y^{(i)}\frac{h_i\sin(t\OM_i)}{\OM_i}
		+ \sg_z^{(i)}\frac{\DE_i h_i[\cos(t\OM_i)-1]}{\OM_i^2},\nonumber\\
		\sg_y^{(i)}(t) &= \sg_x^{(i)}\frac{h_i\sin(t\OM_i)}{\OM_i}
		+ \sg_y^{(i)}\cos(t\OM_i)
		+ \sg_z^{(i)}\frac{\DE_i\sin(t\OM_i)}{\OM_i},\\
		\sg_z^{(i)}(t) &= \sg_x^{(i)}\frac{\DE_i h_i[\cos(t\OM_i)-1]}{\OM_i^2}
		- \sg_y^{(i)}\frac{\DE_i\sin(t\OM_i)}{\OM_i}
		+ \sg_z^{(i)}\frac{\DE_i^2\cos(t\OM_i)+h_i^2}{\OM_i^2},\nonumber
	\end{align}
	\end{widetext}
	where $\OM_i = \sqrt{\DE_i^2+h_i^2}$ is the transition frequency.

	\subsection{Bosonic operators}

	The equations for the bosonic operators are
	\begin{align*}
		\dfrac{\pr}{\pr t}b_n&=-iH_{nn}b_n,\\
		\dfrac{\pr}{\pr t}b_n^\dagger&=iH_{nn}b_n^\dagger,
	\end{align*}
	with the solutions
	\begin{equation}
		\label{eq:BossSol}
	\begin{aligned}
		b_n(t)&=e^{-iH_{nn}t}b_n,\\
		b_n^\dagger(t)&=e^{iH_{nn}t}b_n^\dagger.
	\end{aligned}
	\end{equation}

	\section{Derivation of the equations of motion}\label{ap:B}
	\begin{remark}
		\label{rem:GausAnz}
		Throughout the derivation, we use the Gaussian ansatz $\rho_B=Z_B^{-1}\exp(\mf a^T K\mf a)$ with a diagonal matrix $K$. Consequently, odd-order expectation values and second-order expectation values with $m\neq n$ vanish:
		\begin{equation*}
			\langle b_{m_1}^{\#_1}\cdots b_{m_{2k+1}}^{\#_{2k+1}}\rangle = 0,
			\qquad
			\langle b_m^{\#} b_n^{\#'}\rangle = 0 \quad (m\neq n),
		\end{equation*}
		where $\#_i$ denotes an arbitrary conjugation.
	\end{remark}
	
	Eq.~\eqref{eq:secOrderEqE} then reduces to
	\begin{align*}
		\dfrac{d}{dt}\vec{\mathcal E}(t)
		&=\la^2\Tr\int_{t_0}^t\!dt_1\,\vec P\,
		\mathcal L(t)\mathcal L(t_1)\nonumber\\
		&\quad\times\rho_{\mathrm{ans}}(\vec{\mathcal E}(t)).
	\end{align*}
	Hereafter, a prime on an operator indicates that it is evaluated at time $t_1$, for example, $S_p'=S_p(t_1)$ and $E_p'=E_p(t_1)$.

	\subsection{Equation for the system density matrix}

	For an arbitrary system operator $P_S$, we have
	\begin{align*}
		\dfrac{d}{dt}\braket{P_S}
		&=-\la^2\int_{t_0}^t\!dt_1\sum_{p,p'}
		\Tr\Bigl(P_S[S_p\otimes E_p,\nonumber\\
		&\hspace{20mm}[S_{p'}'\otimes E_{p'}',
		\rho_S\otimes\rho_B]]\Bigr).
	\end{align*}
	Expanding the double commutator and averaging over the environment gives
	\begin{align*}
		&\Tr_B[S_p\otimes E_p,
		[S_{p'}'\otimes E_{p'}',\rho_S\otimes\rho_B]]\\
		&\quad=\bigl(S_pS_{p'}'\rho_S-S_{p'}'\rho_SS_p\bigr)
		C_{pp'}(t,t_1)\\
		&\qquad+\bigl(\rho_SS_{p'}'S_p-S_p\rho_SS_{p'}'\bigr)
		C_{p'p}(t_1,t).
	\end{align*}
	Since this expression holds for any $P_S$, we obtain Eq.~\eqref{eq:RedfForRhoS}.

	\subsection{Equation for the mean spin operator}

	We substitute $P=\sg^{(p)}_\al\otimes I$ into Eq.~\eqref{eq:EqForPComms}:
	\begin{align*}
		&[S_i'\otimes E_i',
		[S_j\otimes E_j,\sg^{(p)}_\al\otimes I]]\\
		&=[S_i'\otimes E_i',
		[S_j,\sg^{(p)}_\al]\otimes E_j]\\
		&=S_i'[S_j,\sg^{(p)}_\al]\otimes E_i'E_j
		-[S_j,\sg^{(p)}_\al]S_i'\otimes E_jE_i'.
	\end{align*}
	We evaluate each factor using $[\sg^{(i)}_k,\sg^{(p)}_\al]=2i\delta_{ip}\,\ve^{k\al\be}\sg^{(p)}_\be$:
	\begin{align*}
		S_i'[S_j,\sg^{(p)}_\al]
		&=\sum_{\be\ga}c'_{\be i}c_{\ga j}\,
		\sg_\be^{(i)}[\sg^{(j)}_\ga,\sg^{(p)}_\al]\\
		&=2i\delta_{jp}\sum_{\be\ga\de}c'_{\be i}c_{\ga j}
		\ve^{\ga\al\de}\sg_\be^{(i)}\sg_\de^{(p)},\\[4pt]
		[S_j,\sg^{(p)}_\al]S_i'
		&=\sum_{\be\ga}c'_{\be i}c_{\ga j}
		[\sg^{(j)}_\ga,\sg^{(p)}_\al]\sg^{(i)}_\be\\
		&=2i\delta_{jp}\sum_{\be\ga\de}c'_{\be i}c_{\ga j}
		\ve^{\ga\al\de}\sg_\de^{(p)}\sg_\be^{(i)}.
	\end{align*}
	Substituting the result into Eq.~\eqref{eq:EqForPComms}, defining $\braket{E_i(t_1)E_p(t)}=C_{ip}(t_1,t)$ and $\braket{E_p(t)E_i(t_1)}=C_{pi}(t,t_1)$, and summing over $i,j$, we obtain
	\begin{align*}
		&2i\sum_i\sum_{\be\ga\de}c'_{\be i}c_{\ga p}
		\ve^{\ga\al\de}\Bigl[
		\sg_{\be}^{(i)}\sg_{\de}^{(p)}C_{ip}(t_1,t)\\
		&\hspace{25mm}-\sg_{\de}^{(p)}\sg_{\be}^{(i)}C_{pi}(t,t_1)\Bigr].
	\end{align*}
	
	Since $[\sg_\al^{(i)},\sg_\be^{(p)}]=0$ for $i\neq p$, we consider the cases $i=p$ and $i\neq p$ separately.
	
	\begin{itemize}
		\item $i = p$
		
		In this case, $\sg_\al\sg_\be=\de^{\al\be}I_2+i\ve^{\al\be\ga}\sg_\ga$, and therefore
		
		\begin{equation*}
		\begin{aligned}
			2i\sum_{\be\ga\de}c'_{\be p}c_{\ga p}\ve^{\ga\al\de}\Bigl[&
			\bigl(\de^{\be\de}I_2+i\ve^{\be\de\nu}\sg_\nu\bigr)C_{pp}(t_1,t)\\
			&-\bigl(\de^{\de\be}I_2+i\ve^{\de\be\nu}\sg_\nu\bigr)C_{pp}(t,t_1)\Bigr].
		\end{aligned}
		\end{equation*}
		
		Collecting the terms proportional to $I_2$ and $\sg_\nu$, we obtain
		
		\begin{equation}
			\label{eq:Parti=p}
		\begin{aligned}
			Q={}&-2\sum_{\be\ga}c_{\be p}(t_1)c_{\ga p}(t)
			\Biggl[\sum_{\de\nu}\ve^{\ga\al\de}\ve^{\be\de\nu}\sg^{(p)}_\nu\\
			&\quad\times\bigl\{C_p(t,t_1)+C_p(t_1,t)\bigr\}\\
			&\quad+i\ve^{\ga\al\be}
			\bigl\{C_p(t,t_1)-C_p(t_1,t)\bigr\}\Biggr].
		\end{aligned}
		\end{equation}
		where $C_p(t,t_1)=C_{pp}(t,t_1)$.
		
		\item $i \not= p$
		
		In this case, the spin operators commute, and we obtain
		
		\begin{align}
			\label{eq:Parti!=p}
			D&=2i\sum_{i\neq p}\sum_{\be\ga\de}
			c_{\be i}(t_1)c_{\ga p}(t)\ve^{\ga\al\de}\nonumber\\
			&\quad\times\sg_\be^{(i)}\sg_\de^{(p)}
			\bigl[C_{ip}(t_1,t)-C_{pi}(t,t_1)\bigr].
		\end{align}
	\end{itemize}
	
	Finally, combining the terms $Q$ and $D$ from Eqs.~\eqref{eq:Parti=p}--\eqref{eq:Parti!=p}, we obtain the equation for $\braket{\sg_\al^{(p)}}$:
	
	\begin{widetext}
	\begin{align}
		\label{eq:EqForSgn}
		\dfrac{d}{dt}\braket{\sg_\al^{(p)}}
		&=2\la^2\int_{t_0}^t\!dt_1\Biggl\{
		\sum_{\be\ga}c_{\be p}(t_1)c_{\ga p}(t)
		\sum_{\de\nu}\ve^{\ga\al\de}\ve^{\be\de\nu}\braket{\sg^{(p)}_\nu}
		\bigl\{C_p(t,t_1)+C_p(t_1,t)\bigr\}\nonumber\\
		&\quad+i\sum_{\be\ga}c_{\be p}(t_1)c_{\ga p}(t)\ve^{\ga\al\be}
		\bigl\{C_p(t,t_1)-C_p(t_1,t)\bigr\}\nonumber\\
		&\quad-i\sum_{i\neq p}\sum_{\be\ga\de}c_{\be i}(t_1)c_{\ga p}(t)\ve^{\ga\al\de}
		\braket{\sg_\be^{(i)}\sg_\de^{(p)}}
		\bigl[C_{ip}(t_1,t)-C_{pi}(t,t_1)\bigr]\Biggr\}.
	\end{align}
	\end{widetext}
	For Hermitian environmental operators, $C_{pi}(t_1,t)=C_{ip}^*(t,t_1)$.

	\subsection{Equation for the {mean mode occupation number}}

	We substitute $P=I_2\otimes N_m$ into Eq.~\eqref{eq:EqForPComms}. Since $\rho_B$ is diagonal, $\braket{[E,N_m]}=0$, and
	\begin{align*}
		&[S_i'\otimes E_i',
		[S_j\otimes E_j,I_2\otimes N_m]]\\
		&=[S_i'\otimes E_i',S_j\otimes[E_j,N_m]]\\
		&=S_i'S_j\otimes E_i'[E_j,N_m]
		-S_jS_i'\otimes[E_j,N_m]E_i'.
	\end{align*}
	
	We consider two cases.
	
	\begin{itemize}
		\item $i = j$
		
		For brevity, we omit the index $i$ on the coefficients and operators.
		
		We first evaluate the products of spin operators:
		\begin{align}
			\label{eq:S'S}
			S'S&=\sum_\al c_\al'c_\al I_2
			+i\sum_{\al\be\ga}c_\al'c_\be
			\ve^{\al\be\ga}\sg_\ga,\nonumber\\
			SS'&=\sum_\al c_\al c_\al'I_2
			+i\sum_{\al\be\ga}c_\al c_\be'
			\ve^{\al\be\ga}\sg_\ga.
		\end{align}	
		
		Since the operators $S$ and $S'$ are Hermitian, $SS'=(S'S)^\dagger$.
		
		We now evaluate the bosonic part:
		\begin{align*}
			E'[E,N_m]
			&=E'(g_mb_m-g_m^*b_m^\dagger)\\
			&=\sum_n(g_n'b_n+g_n'^*b_n^\dagger)\nonumber\\
			&\quad\times(g_mb_m-g_m^*b_m^\dagger).
		\end{align*}
		Upon averaging over the Gaussian state with a diagonal covariance matrix, only the terms $b_n^\dagger b_n$ and $b_nb_n^\dagger$ remain; hence,
		
		\begin{align}
			\label{eq:E'EN}
			\braket{E'[E,N_m]}
			&=g_m'^*g_m\braket{N_m}-g_m'g_m^*(1+\braket{N_m}).
		\end{align} 
		
		Similarly,
		\begin{align}
			\label{eq:ENE'}
			\braket{[E,N_m]E'}
			&=g_mg_m'^*(1+\braket{N_m})-g_m'g_m^*\braket{N_m}.
		\end{align}
		
		Combining Eqs.~\eqref{eq:E'EN} and~\eqref{eq:ENE'}, we obtain
		
		\begin{align*}
			&S'S\bigl[g_m'^*g_m\braket{N_m}-g_m'g_m^*(1+\braket{N_m})\bigr]\\
			&\quad-(S'S)^\dagger
			\bigl[g_m'^*g_m(1+\braket{N_m})-g_m'g_m^*\braket{N_m}\bigr]\\[3pt]
			&=\braket{N_m}\Bigl[S'S(g_m'^*g_m-g_m'g_m^*)\nonumber\\
			&\qquad-(S'S)^\dagger(g_m'^*g_m-g_m'g_m^*)\Bigr]\\
			&\quad-\bigl[S'Sg_m'g_m^*+(S'S)^\dagger g_m'^*g_m\bigr]\\[3pt]
			&=-4\braket{N_m}\Im(S'S)\Im(g_m'^*g_m)\nonumber\\
			&\quad-2\Re(S'Sg_m'g_m^*).
		\end{align*}
		
		Substituting Eq.~\eqref{eq:S'S} and using $g'_mg_m^*=\Re(g'_mg_m^*)+i\Im(g'_mg_m^*)$, we obtain
		
		\begin{align*}
			&-4\braket{N_m}\sum_{\al\be\ga}c_\al'c_\be
			\ve^{\al\be\ga}\sg_\ga\Im(g_m'^*g_m)\\
			&\quad-2\sum_\al c_\al'c_\al\Re(g'_mg^*_m)\\
			&\quad+2\sum_{\al\be\ga}c'_\al c_\be
			\ve^{\al\be\ga}\sg_\ga\Im(g'_mg_m^*).
		\end{align*}
		Since $\Im(ab^*)=-\Im(a^*b)$, collecting like terms gives
		
		\begin{align*}
			&-2(2\braket{N_m}+1)\sum_{\al\be\ga}
			\Im(g_m'^*g_m)c_\al'c_\be
			\ve^{\al\be\ga}\sg_\ga\\
			&\quad-2\sum_\al c'_\al c_\al\Re(g'_mg_m^*).
		\end{align*}
		
		Averaging this expression, we obtain
		\begin{align}
			\label{eq:Ni=j}
			Q&=-\sum_i\Biggl[2(2\braket{N_m}+1)
			\mathrm{Im}\bigl(g^*_{mi}(t_1)g_{mi}(t)\bigr)\nonumber\\
			&\quad\times\sum_{\alpha\beta\gamma}
			c_{\alpha i}(t_1)c_{\beta i}(t)
			\varepsilon^{\alpha\beta\gamma}
			\langle\sigma^{(i)}_\gamma\rangle\nonumber\\
			&\quad+2\sum_\alpha c_{\alpha i}(t_1)c_{\alpha i}(t)
			\mathrm{Re}\bigl(g_{mi}(t_1)g^*_{mi}(t)\bigr)\Biggr]
		\end{align}
		\item $i \not=j$
		
		In this case, the spin operators commute, i.e., $S_i'S_j=S_jS_i'$, which simplifies the expression:
		\begin{align*}
			&S_i'S_j\otimes E_i'[E_j,N_m]
			-S_i'S_j\otimes[E_j,N_m]E_i'\\
			&\qquad=S_i'S_j\otimes[E_i',[E_j,N_m]].
		\end{align*}
		
		The commutator is
		\begin{align*}
			[E_i,N_m]
			&=\left[\sum_n(g_{ni}b_n+g_{ni}^*b_n^\dagger),
			b_m^\dagger b_m\right]\\
			&=g_{mi}b_m-g_{mi}^*b_m^\dagger.
		\end{align*}
		
		The double commutator is therefore
		
		\begin{align*}
			[E_i',[E_j,N_m]]
			&=-(g_{mi}'g_{mj}^*+g_{mi}'^*g_{mj})\\
			&=-2\Re\bigl(g_{mi}'g_{mj}^*\bigr).
		\end{align*}
		
		After averaging, we obtain
		\begin{align}
			\label{eq:forNi!=j}
			D&=-2\sum_{i\neq j}\sum_{\al\be}
			c_{\al i}(t_1)c_{\be j}(t)\nonumber\\
			&\quad\times\braket{\sg_\al^{(i)}\sg_\be^{(j)}}
			\Re\bigl(g_{mi}(t_1)g_{mj}^*(t)\bigr).
		\end{align}
	\end{itemize}
	
	Combining $Q$ and $D$ from Eqs.~\eqref{eq:Ni=j}--\eqref{eq:forNi!=j}, we obtain
	
	\begin{widetext}
	\begin{align}
		\label{eq:EqForNn}
		\dfrac{d}{dt}\braket{N_m}
		&=2\la^2\int_{t_0}^t\!dt_1\Biggl\{\sum_i\Biggl[(2\braket{N_m}+1)\,\mathrm{Im}\bigl(g^*_{mi}(t_1)g_{mi}(t)\bigr)
		\sum_{\alpha\beta\gamma}c_{\alpha i}(t_1)c_{\beta i}(t)\varepsilon^{\alpha\beta\gamma}\braket{\sigma^{(i)}_\gamma}\nonumber\\
		&\qquad+\sum_\alpha c_{\alpha i}(t_1)c_{\alpha i}(t)\mathrm{Re}\bigl(g_{mi}(t_1)g^*_{mi}(t)\bigr)\Biggr]\nonumber\\
		&\qquad+\sum_{i\neq j}\sum_{\al\be}c_{\al i}(t_1)c_{\be j}(t)\braket{\sg_\al^{(i)}\sg_\be^{(j)}}\Re\bigl(g_{mi}(t_1)g_{mj}^*(t)\bigr)\Biggr\}.
	\end{align}
	\end{widetext}

\bibliographystyle{apsrev4-2}
\bibliography{lit_v1.0.6}

@book{Breuer2002,
  author = {Breuer, Heinz-Peter and Petruccione, Francesco},
  title = {The Theory of Open Quantum Systems},
  publisher = {Oxford University Press},
  address = {Oxford},
  year = {2002}
}

@book{Weiss2012,
  author = {Weiss, Ulrich},
  title = {Quantum Dissipative Systems},
  edition = {4},
  publisher = {World Scientific},
  address = {Singapore},
  year = {2012}
}

@article{Zurek2003,
  author = {Zurek, Wojciech H.},
  title = {Decoherence, einselection, and the quantum origins of the classical},
  journal = {Rev. Mod. Phys.},
  volume = {75},
  pages = {715},
  year = {2003},
  doi = {10.1103/RevModPhys.75.715}
}

@article{Redfield1957,
  author = {Redfield, Alfred G.},
  title = {On the Theory of Relaxation Processes},
  journal = {IBM J. Res. Dev.},
  volume = {1},
  pages = {19},
  year = {1957},
  doi = {10.1147/rd.11.0019}
}

@article{Lindblad1976,
  author = {Lindblad, Goran},
  title = {On the generators of quantum dynamical semigroups},
  journal = {Commun. Math. Phys.},
  volume = {48},
  pages = {119},
  year = {1976},
  doi = {10.1007/BF01608499}
}

@article{Gorini1976,
  author = {Gorini, Vittorio and Kossakowski, Andrzej and Sudarshan, E. C. G.},
  title = {Completely positive dynamical semigroups of $N$-level systems},
  journal = {J. Math. Phys.},
  volume = {17},
  pages = {821},
  year = {1976},
  doi = {10.1063/1.522979}
}

@article{deVega2017,
  author = {de Vega, Ines and Alonso, Daniel},
  title = {Dynamics of non-Markovian open quantum systems},
  journal = {Rev. Mod. Phys.},
  volume = {89},
  pages = {015001},
  year = {2017},
  doi = {10.1103/RevModPhys.89.015001}
}

@article{Breuer2016,
  author = {Breuer, Heinz-Peter and Laine, Elsi-Mari and Piilo, Jyrki and Vacchini, Bassano},
  title = {Colloquium: Non-Markovian dynamics in open quantum systems},
  journal = {Rev. Mod. Phys.},
  volume = {88},
  pages = {021002},
  year = {2016},
  doi = {10.1103/RevModPhys.88.021002}
}

@article{Nakajima1958,
  author = {Nakajima, Sadao},
  title = {On Quantum Theory of Transport Phenomena},
  journal = {Prog. Theor. Phys.},
  volume = {20},
  pages = {948},
  year = {1958},
  doi = {10.1143/PTP.20.948}
}

@article{Zwanzig1960,
  author = {Zwanzig, Robert},
  title = {Ensemble Method in the Theory of Irreversibility},
  journal = {J. Chem. Phys.},
  volume = {33},
  pages = {1338},
  year = {1960},
  doi = {10.1063/1.1731409}
}

@article{Shibata1977,
  author = {Shibata, Fumiaki and Takahashi, Yoshinori and Hashitsume, Natsuki},
  title = {A generalized stochastic Liouville equation: Non-Markovian versus memoryless master equations},
  journal = {J. Stat. Phys.},
  volume = {17},
  pages = {171},
  year = {1977},
  doi = {10.1007/BF01040100}
}

@article{Chaturvedi1979,
  author = {Chaturvedi, S. and Shibata, F.},
  title = {Time-convolutionless projection operator formalism for elimination of fast variables: Applications to Brownian motion},
  journal = {Z. Phys. B},
  volume = {35},
  pages = {297},
  year = {1979},
  doi = {10.1007/BF01319852}
}

@article{Tanimura1989,
  author = {Tanimura, Yoshitaka and Kubo, Ryogo},
  title = {Time Evolution of a Quantum System in Contact with a Nearly Gaussian-Markoffian Noise Bath},
  journal = {J. Phys. Soc. Jpn.},
  volume = {58},
  pages = {101},
  year = {1989},
  doi = {10.1143/JPSJ.58.101}
}

@article{Ishizaki2009,
  author = {Ishizaki, Akihito and Fleming, Graham R.},
  title = {Unified treatment of quantum coherent and incoherent hopping dynamics in electronic energy transfer: Reduced hierarchy equation approach},
  journal = {J. Chem. Phys.},
  volume = {130},
  pages = {234111},
  year = {2009},
  doi = {10.1063/1.3155372}
}

@article{Tanimura2020,
  author = {Tanimura, Yoshitaka},
  title = {Numerically ``exact'' approach to open quantum dynamics: The hierarchical equations of motion ({HEOM})},
  journal = {J. Chem. Phys.},
  volume = {153},
  pages = {020901},
  year = {2020},
  doi = {10.1063/5.0011599}
}

@article{Leggett1987,
  author = {Leggett, Anthony J. and Chakravarty, Sudip and Dorsey, Alan T. and Fisher, Matthew P. A. and Garg, Anupam and Zwerger, Wolfgang},
  title = {Dynamics of the dissipative two-state system},
  journal = {Rev. Mod. Phys.},
  volume = {59},
  pages = {1},
  year = {1987},
  doi = {10.1103/RevModPhys.59.1}
}

@article{Caldeira1983,
  author = {Caldeira, A. O. and Leggett, A. J.},
  title = {Quantum tunnelling in a dissipative system},
  journal = {Ann. Phys.},
  volume = {149},
  pages = {374},
  year = {1983},
  doi = {10.1016/0003-4916(83)90202-6}
}

@book{Nitzan2006,
  author = {Nitzan, Abraham},
  title = {Chemical Dynamics in Condensed Phases},
  publisher = {Oxford University Press},
  address = {Oxford},
  year = {2006}
}

@article{Krantz2019,
  author = {Krantz, Philip and Kjaergaard, Morten and Yan, Fei and Orlando, Terry P. and Gustavsson, Simon and Oliver, William D.},
  title = {A quantum engineer's guide to superconducting qubits},
  journal = {Appl. Phys. Rev.},
  volume = {6},
  pages = {021318},
  year = {2019},
  doi = {10.1063/1.5089550}
}

@article{Wendin2017,
  author = {Wendin, Goran},
  title = {Quantum information processing with superconducting circuits: a review},
  journal = {Rep. Prog. Phys.},
  volume = {80},
  pages = {106001},
  year = {2017},
  doi = {10.1088/1361-6633/aa7e1a}
}

@article{Garraway1997,
  author = {Garraway, B. M.},
  title = {Nonperturbative decay of an atomic system in a cavity},
  journal = {Phys. Rev. A},
  volume = {55},
  pages = {2290},
  year = {1997},
  doi = {10.1103/PhysRevA.55.2290}
}

@article{JaynesCummings1963,
  author = {Jaynes, Edwin T. and Cummings, Frederick W.},
  title = {Comparison of quantum and semiclassical radiation theories with application to the beam maser},
  journal = {Proc. IEEE},
  volume = {51},
  pages = {89},
  year = {1963},
  doi = {10.1109/PROC.1963.1664}
}

@article{Vinjanampathy2016,
  author = {Vinjanampathy, Sai and Anders, Janet},
  title = {Quantum thermodynamics},
  journal = {Contemp. Phys.},
  volume = {57},
  pages = {545},
  year = {2016},
  doi = {10.1080/00107514.2016.1201896}
}

@incollection{Alicki2018,
  author = {Alicki, Robert and Kosloff, Ronnie},
  title = {Introduction to Quantum Thermodynamics: History and Prospects},
  booktitle = {Thermodynamics in the Quantum Regime},
  editor = {Binder, Felix and Correa, Luis A. and Gogolin, Christian and Anders, Janet and Adesso, Gerardo},
  publisher = {Springer},
  address = {Cham},
  pages = {1--33},
  year = {2018},
  doi = {10.1007/978-3-319-99046-0_1}
}

@article{portugal2022effective,
  author = {Portugal, Pedro and Brange, Fredrik and Flindt, Christian},
  title = {Effective temperature pulses in open quantum systems},
  journal = {Phys. Rev. Research},
  volume = {4},
  pages = {043112},
  year = {2022},
  doi = {10.1103/PhysRevResearch.4.043112}
}

@article{kundu2021time,
  author = {Kundu, Sohang and Makri, Nancy},
  title = {Time Evolution of Bath Properties in Spin-Boson Dynamics},
  journal = {J. Phys. Chem. B},
  volume = {125},
  pages = {8137},
  year = {2021},
  doi = {10.1021/acs.jpcb.1c03861}
}

@article{zhang2010effect,
  author = {Zhang, P. and Wu, M. W.},
  title = {Effect of nonequilibrium phonons on hot-electron spin relaxation in n-type {GaAs} quantum wells},
  journal = {EPL},
  volume = {92},
  pages = {47009},
  year = {2010},
  doi = {10.1209/0295-5075/92/47009}
}

@article{link2022non,
  author = {Link, Valentin and Strunz, Walter T. and Luoma, Kimmo},
  title = {Non-Markovian Quantum Dynamics in a Squeezed Reservoir},
  journal = {Entropy},
  volume = {24},
  pages = {352},
  year = {2022},
  doi = {10.3390/e24030352}
}

@article{sukharnikov2026non,
  author = {Sukharnikov, V. and Chuchurka, S. and Schlawin, F.},
  title = {Non-Markovian dynamics in nonstationary Gaussian baths: A hierarchy of pure states approach},
  journal = {Phys. Rev. Research},
  volume = {8},
  pages = {013123},
  year = {2026},
  doi = {10.1103/YT37-S9HZ}
}

@article{menczel2024non,
  author = {Menczel, Paul and Funo, Ken and Cirio, Mauro and Lambert, Neill and Nori, Franco},
  title = {Non-Hermitian pseudomodes for strongly coupled open quantum systems: Unravelings, correlations, and thermodynamics},
  journal = {Phys. Rev. Research},
  volume = {6},
  pages = {033237},
  year = {2024},
  doi = {10.1103/PhysRevResearch.6.033237}
}

@article{meretukov2024time,
  author = {Meretukov, Kh. Sh. and Teretenkov, A. E.},
  title = {On Time-Dependent Projectors and a Generalization of the Thermodynamical Approach in the Theory of Open Quantum Systems},
  journal = {Proc. Steklov Inst. Math.},
  volume = {324},
  pages = {135--152},
  year = {2024},
  doi = {10.1134/S0081543824010140}
}

@article{Kawasaki1973,
  author = {Kawasaki, Kyozi and Gunton, James D.},
  title = {Theory of Nonlinear Transport Processes: Nonlinear Shear Viscosity and Normal Stress Effects},
  journal = {Phys. Rev. A},
  volume = {8},
  pages = {2048},
  year = {1973},
  doi = {10.1103/PhysRevA.8.2048}
}

@article{Robertson1966,
  author = {Robertson, Bruce},
  title = {Equations of Motion in Nonequilibrium Statistical Mechanics},
  journal = {Phys. Rev.},
  volume = {144},
  pages = {151},
  year = {1966},
  doi = {10.1103/PhysRev.144.151}
}

@book{Grabert1982,
  author = {Grabert, Hermann},
  title = {Projection Operator Techniques in Nonequilibrium Statistical Mechanics},
  series = {Springer Tracts in Modern Physics},
  volume = {95},
  publisher = {Springer},
  address = {Berlin},
  year = {1982}
}

@article{Lambert2023,
  author = {Lambert, Neill and Raheja, Tarun and Cross, Simon and Menczel, Paul and Ahmed, Shahnawaz and Pitchford, Ahsan and Burgarth, Daniel and Nori, Franco},
  title = {{QuTiP-BoFiN}: A bosonic and fermionic numerical hierarchical-equations-of-motion library with applications in light-harvesting, quantum control, and single-molecule electronics},
  journal = {Phys. Rev. Research},
  volume = {5},
  pages = {013181},
  year = {2023},
  doi = {10.1103/PhysRevResearch.5.013181}
}

@misc{Meretukov2026Code,
  author = {Meretukov, Khazret},
  title = {{Meretukov-KS/spin-boson-projection-tcl2}: TCL2 source code},
  howpublished = {GitHub repository},
  year = {2026},
  note = {Commit 4dea2400d5b2218a96039ee80d89ff3ece96fefe},
  url = {https://github.com/Meretukov-KS/spin-boson-projection-tcl2/tree/4dea2400d5b2218a96039ee80d89ff3ece96fefe}
}

@article{Gribben2022,
  author = {Gribben, Dominic and Strathearn, Aidan and Fux, Gerald E. and Kirton, Peter and Lovett, Brendon W.},
  title = {Using the Environment to Understand non-Markovian Open Quantum Systems},
  journal = {Quantum},
  volume = {6},
  pages = {847},
  year = {2022},
  doi = {10.22331/q-2022-10-25-847}
}

@article{Lambert2019,
  author = {Lambert, Neill and Ahmed, Shahnawaz and Cirio, Mauro and Nori, Franco},
  title = {Modelling the ultra-strongly coupled spin-boson model with unphysical modes},
  journal = {Nature Communications},
  volume = {10},
  pages = {3721},
  year = {2019},
  doi = {10.1038/s41467-019-11656-1}
}

@article{Chen2023,
  author = {Chen, Lipeng and Yan, Yiying and Gelin, Maxim F. and L{\"u}, Zhiguo},
  title = {Dynamics of the spin-boson model: The effect of bath initial conditions},
  journal = {Journal of Chemical Physics},
  volume = {158},
  pages = {104109},
  year = {2023},
  doi = {10.1063/5.0138399}
}

@article{Thorwart2004,
  author = {Thorwart, Michael and Paladino, Elisabetta and Grifoni, Milena},
  title = {Dynamics of the spin-boson model with a structured environment},
  journal = {Chemical Physics},
  volume = {296},
  pages = {333--344},
  year = {2004},
  doi = {10.1016/j.chemphys.2003.10.007}
}

@article{Sun2025,
  author = {Sun, Ke and Kang, Mingyu and Nuomin, Hanggai and Schwartz, George and Beratan, David N. and Brown, Kenneth R. and Kim, Jungsang},
  title = {Quantum simulation of spin-boson models with structured bath},
  journal = {Nature Communications},
  volume = {16},
  pages = {4042},
  year = {2025},
  doi = {10.1038/s41467-025-59296-y}
}

@article{So2026,
  author = {So, Visal and Zhu, Mingjian and {Duraisamy Suganthi}, Midhuna and Menon, Abhishek and Tomaras, George and Zhuravel, Roman and Pu, Han and Pagano, Guido},
  title = {Experimental Realization of Thermal Reservoirs with Tunable Temperature in a Trapped-Ion Spin-Boson Simulator},
  journal = {Physical Review Letters},
  volume = {137},
  pages = {050604},
  year = {2026},
  doi = {10.1103/MF8D-91WS}
}

@article{Magazzu2018,
  author = {Magazz{\`u}, Luca and Forn-D{\'i}az, Pol and Belyansky, Ron and Orgiazzi, Jean-Luc and Yurtalan, M. A. and Otto, M. R. and Lupascu, Adrian and Wilson, C. M. and Grifoni, Milena},
  title = {Probing the strongly driven spin-boson model in a superconducting quantum circuit},
  journal = {Nature Communications},
  volume = {9},
  pages = {1403},
  year = {2018},
  doi = {10.1038/s41467-018-03626-w}
}

@article{Kitzman2023,
  author = {Kitzman, J. M. and Lane, J. R. and Undershute, C. and Harrington, P. M. and Beysengulov, N. R. and Mikolas, C. A. and Murch, K. W. and Pollanen, J.},
  title = {Phononic bath engineering of a superconducting qubit},
  journal = {Nature Communications},
  volume = {14},
  pages = {3910},
  year = {2023},
  doi = {10.1038/s41467-023-39682-0}
}

@article{TrubilkoBasharov2020,
  author = {Trubilko, A. I. and Basharov, A. M.},
  title = {Hierarchy of Times of Open Optical Quantum Systems and the Role of the Effective Hamiltonian in the White Noise Approximation},
  journal = {JETP Letters},
  volume = {111},
  pages = {532--538},
  year = {2020},
  doi = {10.1134/S0021364020090106}
}

@article{Trushechkin2021,
  author = {Trushechkin, Anton},
  title = {Unified Gorini--Kossakowski--Lindblad--Sudarshan Quantum Master Equation Beyond the Secular Approximation},
  journal = {Physical Review A},
  volume = {103},
  pages = {062226},
  year = {2021},
  doi = {10.1103/PhysRevA.103.062226},
  eprint = {2103.12042},
  archivePrefix = {arXiv},
  primaryClass = {quant-ph}
}

@article{Fruchtman2016,
  author = {Fruchtman, Amir and Lambert, Neill and Gauger, Erik M.},
  title = {When do perturbative approaches accurately capture the dynamics of complex quantum systems?},
  journal = {Scientific Reports},
  volume = {6},
  pages = {28204},
  year = {2016},
  doi = {10.1038/srep28204}
}

\end{document}